\documentclass{aa} 

\usepackage{graphicx}
\usepackage{natbib} 
\usepackage{txfonts} 
\begin{document}

   \title{Data on 824 fireballs observed by the digital cameras \\ of the European Fireball Network 
   in 2017--2018\thanks{The catalog is available at the CDS via ...}}
   \subtitle{ I. Description of the network, data reduction procedures, and the catalog}

   \author{J.~Borovi\v{c}ka \inst{1} \and P.~Spurn\'y\inst{1} \and L.~Shrben\'y\inst{1} \and R.~\v{S}tork\inst{1} \and
    L.~Kotkov\'a \inst{1}\and J.~Fuchs\inst{1} \and J.~Kecl\'{\i}kov\'a \inst{1}
   \and H.~Zichov\'a \inst{1} \and J.~M\'anek\inst{1} \and P.~V\'achov\'a \inst{1}\and I.~Macourkov\'a \inst{1}
   \and J.~Svore\v{n} \inst{2} \and H.~Mucke \inst{3}\thanks{deceased}
          }

   \institute{Astronomical Institute of the Czech Academy of Sciences, Fri\v{c}ova 298, 25165 Ond\v{r}ejov, Czech Republic \\
              \email{jiri.borovicka@asu.cas.cz}
    \and          
           Astronomical Institute of the Slovak Academy of Sciences, 05960 Tatranská Lomnica, Slovak Republic
    \and
         Astronomisches B\"{u}ro, 1230 Wien, Austria 
      }

   \date{Received 7 June 2022, Accepted 25 August 2022 }
   \titlerunning{Data on 824 fireballs observed by the European Fireball Network I}
   \authorrunning{Borovi\v{c}ka et al.}

 
\abstract{A catalog of 824 fireballs (bright meteors), observed by a dedicated network
of all-sky digital photographic cameras in central Europe in the years 2017--2018 is presented. 
The status of the European Fireball Network, established in 1963, is described.
The cameras collect digital images of meteors brighter than an absolute magnitude of about $-2$ and radiometric light curves with a
high temporal resolution of those brighter than a magnitude $\approx -4$. All meteoroids larger than 5 grams, corresponding to
sizes of about 2 cm, are detected regardless
of their entry velocity. High-velocity meteoroids are detected down to masses of about 0.1 gram. The largest observed
meteoroid in the reported period 2017--2018 had a mass of about 100~kg and a size of about 40 cm. The methods of data 
analysis are explained and all catalog entries are described in detail. The provided data include the fireball date and time, 
atmospheric trajectory and velocity, the radiant in various coordinate systems, heliocentric orbital elements, maximum brightness,
radiated energy, initial and terminal masses, maximum encountered dynamic pressure, physical classification, and possible shower
membership. Basic information on the fireball spectrum is available for some bright fireballs (apparent magnitude $< -7$). 
A simple statistical evaluation of the whole 
sample is provided. The scientific analysis is presented in an accompanying paper.}
  


   \maketitle
%

\section{Introduction}

Comets, asteroids, and other bodies of the Solar System can lose solid mass by various
mechanisms \citep[see e.g.,][]{Jewitt}.
Their fragments, dust particles, and meteoroids are subject to both
gravitational and nongravitational forces \citep{VauM19} and evolve independently on the parent body. 
Meteoroids are eventually either expelled from the Solar System;
collide with the Sun, planets, or other bodies; or are destroyed by mutual collisions \citep{Koschny}.
Unless moving together in large quantities and forming diffuse orbital structures, such
as zodiacal dust bands \citep{Nesvorny_dust} or cometary debris trails \citep{Reach}, 
meteoroids cannot be 
observed telescopically in interplanetary space because of their faintness. They
demonstrate themselves most easily during the final stages of their existence when colliding
with planets or other bodies. When penetrating through planetary atmospheres, they form
luminous events called meteors \citep{SSR}. When colliding with airless bodies, they produce impact flashes 
and craters \citep{flashes}.

Meteoroids have wide range of sizes, from $\sim$10$^{-5}$~m to 1~m (larger bodies 
are referred to as asteroids, and smaller ones as dust particles). 
They carry information about the structure, physical properties,
and composition of their parent bodies on these scales. Such information is difficult to get
directly from asteroids and comets since in situ exploration is always restricted to
specific bodies because of logistic and financial limitations. 
Investigation of meteoroids represents an opportunity to sample 
different populations of asteroids and comets.
The most practical way is to use a terrestrial atmosphere 
as a meteoroid detector and observe meteors from the ground. Although such observations have
been carried out since ancient times \citep[e.g.,][]{Hasegawa}, the challenge is to have sufficiently precise
techniques and methods to derive reliable preatmospheric orbits and, which is even more difficult,
to evaluate the physical properties of meteoroids using the short period of time when meteoroids are decaying
in the atmosphere. In rare cases, meteoroid fragments reaching the ground are recovered as meteorites 
and can be investigated in laboratory in detail. However, they represent only the strongest population of meteoroids.

Meteoroids of different sizes need different techniques. The smallest meteoroids are numerous, but
produce faint meteors. They are best observed by radars, which detect the produced ionization 
\citep{Kerochapter}. Brighter meteors are nowadays widely observed by video cameras. 
In this paper, we are mostly interested in centimeter-sized and larger meteoroids. They produce bright meteors
(fireballs), which are relatively rare. In order to obtain enough data, long-term monitoring of a large
volume of the    atmosphere is needed. Fireball networks consisting of multiple stations equipped
with wide-angle or all-sky cameras of moderate sensitivity observing every clear night serve for
this purpose. 

Here we present fireball data from two years of observation by the digital photographic cameras of the
European Fireball Network. In contrast to radar and video systems which produce millions and hundreds of
thousands of meteor orbits, respectively, obtained by automated software procedures \citep[e.g.,][]{CMOR,CAMS}, 
our sample is less
numerous, but every single meteor has been measured and reduced by partly manual and partly semi-automatic procedures
with the aim to obtain
the best possible accuracy of the data. The precision and accuracy of the cameras and 
the procedures used has been demonstrated by the identification of the resonant structure within
the Taurid meteoroid stream \citep{Spurny_Taurids}. The advantage of the network is that most fireballs
are detected by more than two cameras, enabling a cross-check.

Other all-sky camera networks aimed at bright fireballs have been build in recent years. 
The Fireball Recovery and InterPlanetary Observation Network (FRIPON) \citep{FRIPON}
 has larger geographic coverage, but uses all-sky video cameras
of relatively low resolution and the data are inevitably of a lower precision. The Desert Fireball Network (DFN) 
has grown from its initial stage \citep{Bland_DFN} into a much larger network and now uses 
 digital photographic cameras similar to ours \citep{Howie}. The comparison of Taurid data contained
in \citet{Spurny_Taurids} and \citet{DFN_Taurids}, in particular of the semimajor axes of the resonant branch,
nevertheless, shows that DFN accuracy is somewhat lower, probably because of automated
reduction procedures and/or a lower quality of the optics used. Both of these networks also have problems with 
photometry, especially for brighter fireballs, since radiometers are not used. The privately built and rapidly
growing All-Sky-7 network \citep[formerly All-Sky-6,][]{AllSky6} uses seven
video cameras at each station, which together cover the whole sky, and the network has good potential 
but is more suited to fainter meteors. Other networks, for example 
those built in Spain \citep{SPMN}, Canada \citep{SOMN}, the USA \citep{NASAallsky}, or Poland \citep{PFN}, mostly use
low-resolution video cameras.

The data presented here include the atmospheric trajectory and velocity, radiant, heliocentric orbit,
maximum brightness, total radiated energy, and maximum encountered dynamic pressure. 
An estimate of the meteoroid initial mass based on fireball radiation and velocity 
is provided. The possible shower membership and possible parent body are listed. In addition to
the errors of radiant and velocity, supplementary quality criteria such as the number of cameras used
and the minimal distance between the closest camera and the fireball are given. 
Physical properties of meteoroids
are evaluated according to the classical PE criterion
based on the fireball end height \citep{PE}\footnote{The meaning of the PE acronym was not clearly
 explained, but was probably derived from the graphical similarity to $\rho_{\rm E}$ designing the atmospheric
 density at the fireball end.}
and according to the newly proposed pressure factor based on the maximum dynamic pressure.
Basic information on the fireball spectrum, if available, is also provided.

The data form a catalog provided at the  \textit{Centre de Donn\'ees astronomiques de Strasbourg} (CDS).
Each fireball has one (long)
row in the catalog. This work does not contain detailed light curves or deceleration data; although, they
are available for the vast majority of the fireballs. Such data can be used for modeling meteoroid
atmospheric fragmentation, which was in fact done for some bright and deeply penetrating fireballs
from this sample in \citet{2strengths}. The light curve and deceleration data may be published 
in the future.

Section \ref{network} provides the overview of the camera network and the description of digital cameras. 
The procedures used for image measurements and the computation of the fireball trajectory, velocity, light curve, and orbit
are described in Section \ref{reduction}. The estimation of the meteoroid mass and its physical classification
using the PE criterion are also explained there. Section~\ref{catalog} is devoted to the explanation of individual
catalog entries. Finally, the summary description and some statistics of the whole sample is given 
in Section~\ref{description}. This section is supplemented by histograms in Appendix~\ref{histograms}.

The data presented here were further used to study physical and orbital properties of centimeter-sized meteoroids in the Solar
System. The analysis, dealing with both sporadic meteoroids and meteoroid streams, is presented in
the accompanying paper \citep[][hereafter Paper II]{paper2}.

   \begin{figure}
   \centering
   \includegraphics[width=8.5cm]{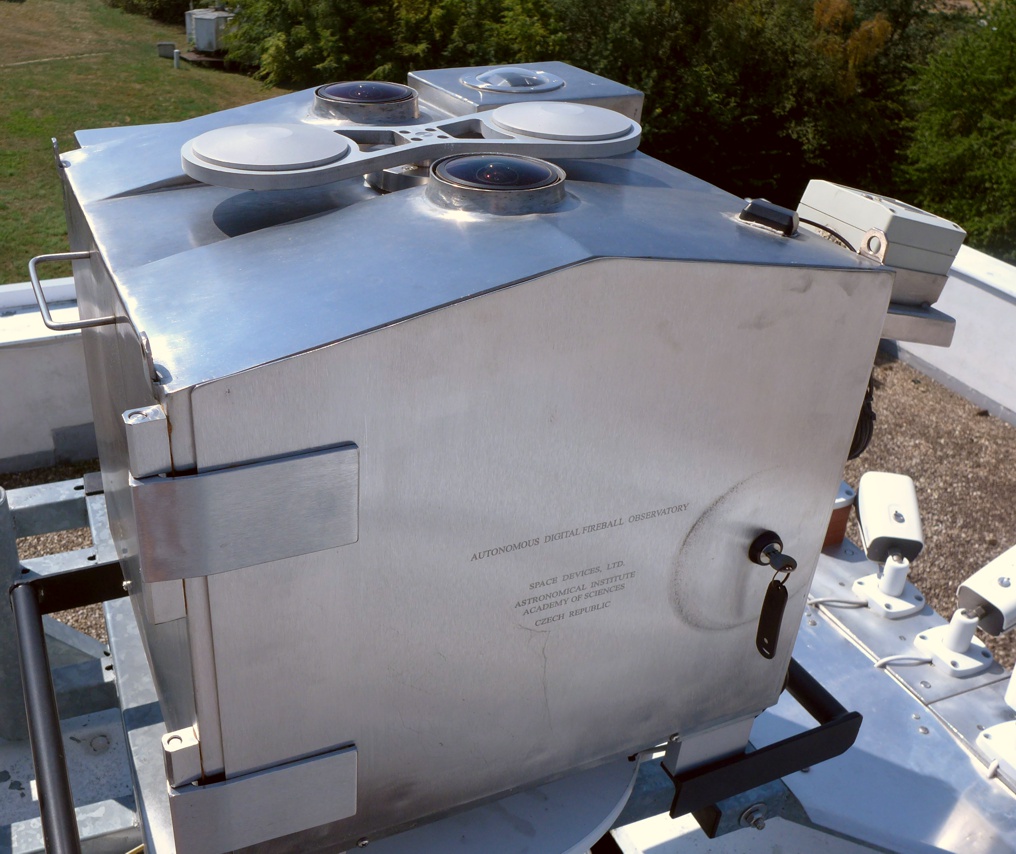}
   \caption{Digital Autonomous Fireball Observatory (DAFO) with the lens cover open at its site in Ond\v{r}ejov.
   The Aurora weather sensor is on the right, next to the GPS receiver. The radiometer window is at the back right corner.}
   \label{DAFO}
   \end{figure}
   
\section{Digital cameras of the European Fireball Network}
\label{network}

\subsection{History of the network}

The first fireball network in the world was established in former Czechoslovakia in 1963 
\citep{CepRaj65}. In the following years, further stations were set up in southern Germany and since 1968
the network has become known as the European Fireball Network \citep{Cep73}. The German part was
gradually extended to other parts of Germany and even outside German borders, but the stations remained 
equipped with low-resolution all-sky mirror cameras \citep{Oberst, Flohrer}. The headquarters of the network
have remained at the Ond\v{r}ejov Observatory, Czech Republic. The Czech part of the network
was modernized several times. In the second half of the 1970s, mirror cameras were replaced by fish-eye cameras 
with higher resolution \citep{bolidy77}. In the first years of the new century, the cameras were automated
and equipped with photoelectric radiometers, creating so-called Autonomous Fireball Observatories
\citep[AFOs;][]{Spurny_IAU}. Finally, from 2013--2015, a new type of Digital Autonomous Fireball Observatory (DAFO)
was developed and deployed \citep{Spurny_Taurids}. The number of stations was also enlarged.

Tables with data of selected fireballs were published in a number of papers 
\citep[e.g.,][]{Cep77, bolidy77, bolidy78, Spurny94, Spurny97, Spurny_Ori, Spurny_Taurids}.
The European Fireball Network also enabled the recovery of several meteorites, for example\ 
Bene\v{s}ov \citep{Benesov}, Neuschwanstein \citep{Neuschw}, and \v{Z}\v{d}\'{a}r nad S\'{a}zavou
\citep{Zdar}, and provided their heliocentric orbits.

\subsection{Digital cameras (DAFOs)}
\label{DAFO}

The installation of digital cameras (DAFOs) began in 2013 and significantly
enhanced the capabilities and efficiency of the network. The development of DAFO was enabled 
by the advancement of digital photographic techniques, the progress in computer technology 
(speed and data storage capacity), and the improvement in the internet coverage of the country. 
DAFO (Fig.~\ref{DAFO}) is a weather-proof, fully autonomous box with a size of about 
$45\times45\times40$ cm with some external accessories. DAFO only needs a power supply to operate. 
Internet connection is used for remote checks and configuration as well as remote data downloads. 
The function of DAFO is to image the whole sky continuously during the night under favorable weather conditions
and to carry out continuous high-frequency monitoring of the total brightness of the sky during the night under 
any weather condition.

The imaging is performed by a pair of commercial digital single lens reflex (DSLR) cameras, including Canon EOS 6D 
equipped with the Sigma 8mm F3.5 EX DG Circular Fish-eye lenses. The complementary metal–oxide–semiconductor (CMOS) 
sensor has a resolution
of $5472 \times 3648$ pixels, 14 bits dynamic range, and a physical size  of $35.9\times23.9$ mm. 
The field of view is a full 180\degr; 
the diameter of the sky on the image is about 3500 pixels. Two DSLR cameras are used to avoid any dead time.
Normally, each camera takes exposures 35 seconds long with breaks of 25 seconds. The starts of the exposures
of both cameras are shifted by 30 seconds. There are therefore 5 seconds of overlap every 30 seconds, 
where both cameras are working. It is ensured this way that any fireball shorter than 5 seconds is completely 
imaged by at least one camera. For longer fireballs, it may happen that images from both cameras must be combined.
In the case of the failure of one of the cameras (mainly due to a failure of the shutter, which has a limited lifetime),
DAFO can be configured to a one-camera operation. In that case, one camera is taking exposures continuously
with gaps of about 1.5 seconds between images. The gaps are needed to read out and save the image. 
The exposure length of 35 seconds was chosen so that stars remain point-like and easy to measure on the whole
image.

The lenses are protected by a mechanical cover. During the exposure, the cover is open (see Fig.~\ref{DAFO})
and the lenses are directed to the sky without any other optical element (e.g., a transparent dome). 
The full aperture (f/3.5) is used. The best focusing was found by trial and error for each camera before the DAFO
was assembled and it has since been stable during the camera lifetime.

To measure the fireball speed, a liquid crystal display (LCD) shutter was added on the back side of the lens. This
shutter replaces the mechanical rotating shutter used in the AFO. The X-FOS(G2)-CE 
(Extra Fast Optical Shutter, 2nd generation-Contrast Enhanced) product from the company LC-Tec 
\citep[see also][]{Bettonvil} was used. The closing time is $\leq 50$ $\mu$s and the opening time is 
$\leq 1.8$ ms. The transmittance is $\sim$30\% in the open state and virtually zero in the closed
state, with a contrast $>$ 1:10$^5$. The control electronics ensures alternation between the open and closed
states in a 1:1 ratio during the exposure with exact frequency. One from ten prescribed frequencies 
between 10 Hz and 60 Hz can be selected for each night. Usually, the frequency of 16 Hz is used.
That would mean 16 open and 16 closed states during each second, in other words\ the moving fireball would produce
16 dashes, hereafter called shutter breaks, per second. However, the first closed state during each second
was skipped intentionally to produce a time mark, that is\ a three times longer shutter break (something not possible
with mechanical shutter).  In combination with the fireball light curve from the radiometer, the time mark enabled
us to determine the absolute time of each shutter break with millisecond precision. 
All parameters can be changed remotely.

 \begin{table}
\caption{List of stations equipped with DAFO as of 2018}
\label{stationlist}
\begin{small}
\begin{tabular}{r@{\hspace{2mm}} l c c c l l}
\hline  \noalign{\smallskip}
No. & Name & $\lambda$  & $\varphi$ & $h$ & Start & DAFO  \\
   && \degr E & \degr N & m  & year & start date \\
\hline  \noalign{\smallskip}
 1& \v Sindelov\'a    & 12.597 & 50.317 &  595 & 2015 & 2015-06-30 \\
 2& Kun\v zak       & 15.201 & 49.107 &  656 & 2005 & 2013-12-20 \\
 3& R\r u\v zov\' a       & 14.287 & 50.834 &  348 & 1995 & 2014-05-14 \\
 4& Chur\'a\v nov     & 13.615 & 49.068 & 1119 & 1964 & 2014-02-04 \\
 5& Kocelovice   & 13.838 & 49.467 &  525 & 2015 & 2015-07-01 \\
 6& Fr\'ydlant     & 15.090 & 50.918 &  345 & 2017 & 2017-09-12 \\
 7& Kucha\v rovice  & 16.086 & 48.881 &  340 & 2009 & 2014-12-09 \\
 9& Svratouch    & 16.034 & 49.735 &  736 & 1963 & 2014-11-25 \\
10& Polom        & 16.322 & 50.350 &  748 & 2005 & 2014-11-12 \\
11& P\v rimda       & 12.678 & 49.669 &  745 & 1981 & 2014-06-25 \\
12& Vesel\'\i\ nad    & 17.370 & 48.954 &  176 & 1964 & 2014-12-04 \\[-0.25ex]
    & Moravou \\
14& \v Cerven\'a hora      & 17.542 & 49.777 &  749 & 1977 & 2014-11-27 \\
16& Lys\'a hora    & 18.448 & 49.546 & 1324 & 1991 & 2013-11-13 \\
20& Ond\v rejov     & 14.780 & 49.910 &  527 & 1963 & 2012-06\tablefootmark{a}    \\
23& Star\'a Lesn\'a  & 20.288 & 49.152 &  833 & 2009 & 2014-12-11 \\
24& Kolonica     & 22.274 & 48.935 &  456 & 2016 & 2016-11-10 \\
25& Rimavsk\'a   & 20.005 & 48.374 &  230 & 2018 & 2018-06-26 \\
    & Sobota \\
26& Martinsberg  & 15.126 & 48.381 &  872 & 2009 & 2015-09-24 \\
\hline
\end{tabular}
\end{small}
\tablefoot{
Longitudes and latitudes are given here to three decimal degrees only,
but they were measured with meter precision.\\
\tablefoottext{a}{Start of noncontinuous test operation.}
}
\end{table}

   \begin{figure}
   \centering
   \includegraphics[width=\columnwidth]{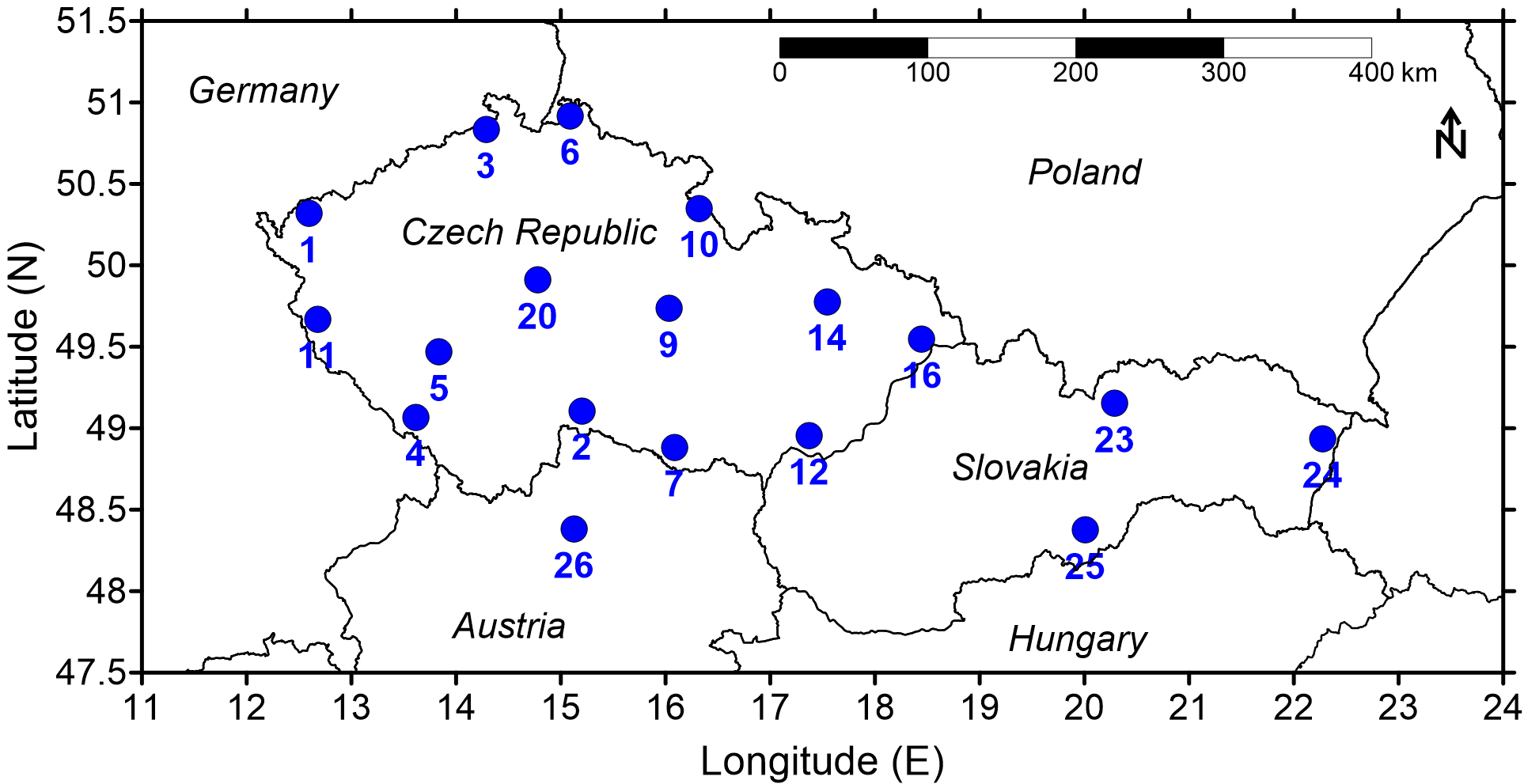}
   \caption{Map of locations of EN fireball stations equipped with DAFO. Valid for the end of 2018.}
   \label{stationmap}
   
   \end{figure}
   
    \begin{figure*}
   \centering
   \includegraphics[width=1.5\columnwidth]{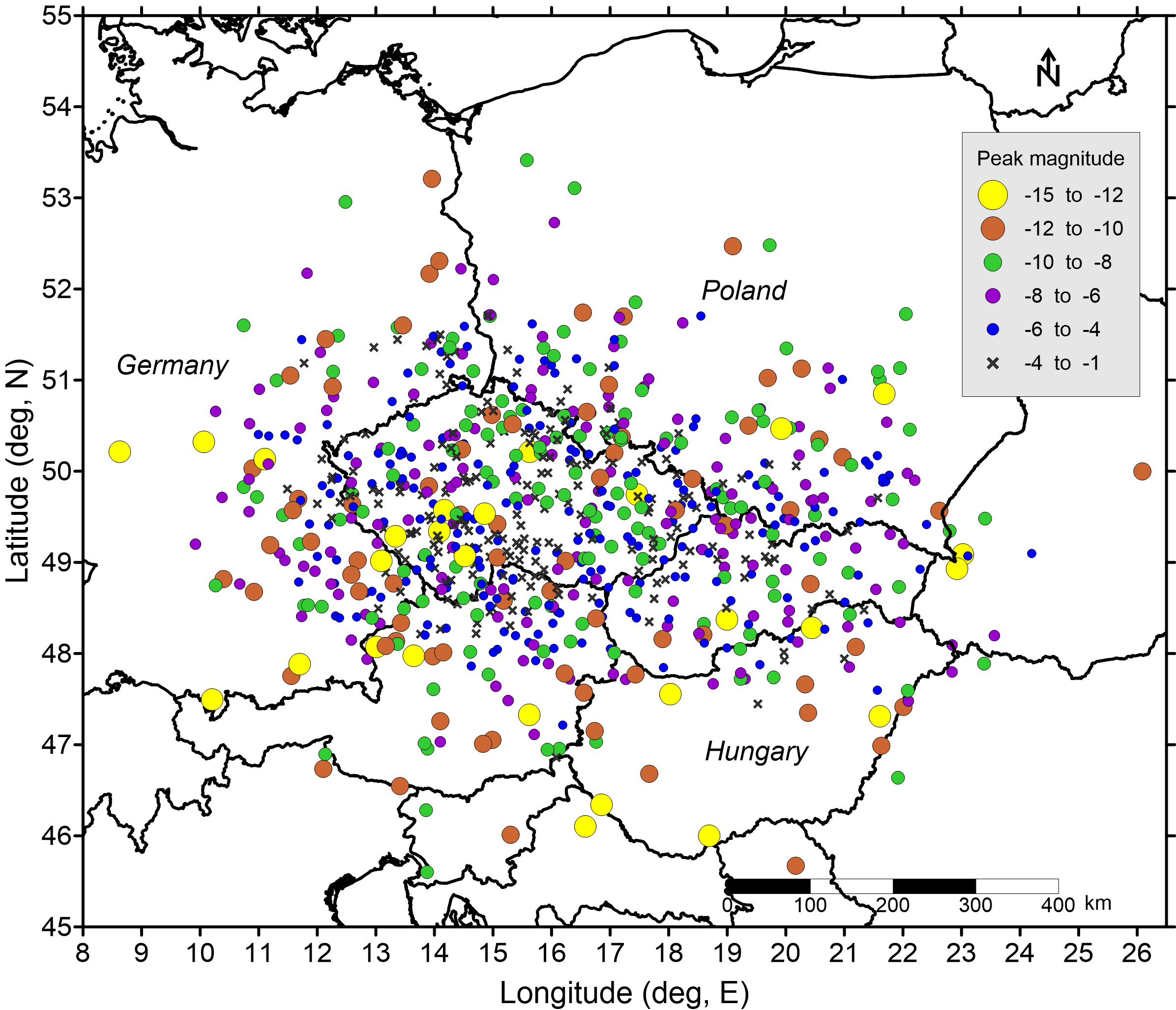}
   \caption{Location of  822 fireballs with measured photometry on the map of central Europe. 
   The plotted positions correspond to the points of maximum brightness. Fireballs are divided into six
   groups according to their maximum brightness.}
   \label{map}
   \end{figure*}  
   
The radiometer monitoring sky brightness is based on a photomultiplier (manufactured by the company Tesla V\'UVET or ET Enterprises) with
an active diameter of 25 mm directed vertically toward the sky without any optics. It monitors the integral brightness of
the sky during the whole night with a sampling frequency of 5000 Hz. The radiometer was placed under a glass
window and it operates during any weather condition since the radiation of bright fireballs can also be detected through a 
cloud cover, even when it is raining. 
During the day, the radiometer is protected by a mechanical shade against direct sunlight.
The control electronics ensures linearity of the output signal in the range of 20 bits, corresponding to the
range of input light intensities of 1:$10^6$. Self-regulation of the high voltage ensures that the output signal
is kept on a constant level (usually 1000 units) unless a rapid change in sky brightness (e.g., due to a fireball or lightning)
occurs. The signal is monitored in real time by the control computer. In the case of a detected event, the data are saved
separately in full time resolution (5000 Hz). Events brighter than a certain limit are reported immediately by email.
Whole night data are saved in reduced resolution (500 Hz).
   
Both the radiometer and the LCD shutter are supplied with a precise time signal from the Precise Positioning Service (PPS)
of the Global Positioning System (GPS) receiver. The absolute time is therefore known for any data point
from the radiometer. Using the time marks on fireball images from the cameras, the radiometric light curve
can be connected with the fireball trajectory, that is\ both the fireball position and brightness are known as a function
of absolute time.

The weather is monitored by the cloud sensor by Aurora Eurotech, which also detects precipitation. 
The lens cover is open and exposure is taken only in periods without precipitation and when the sky
is at least partly clear. DAFO housing is heated to melt any fallen snow.
Temperatures inside DAFO are kept in a reasonable working range by internal heating
or by cooling using a fan on the bottom of DAFO. Some DAFOs also have an active cooling of the cameras based on
Peltier modules. Thermal noise was found to be a serious problem during the summer in the images taken with the
previous version of cameras (Canon EOS 5D Mark II). Canon 6D, now used in all DAFOs, can be operated without active 
cooling, especially at sites with a climate that is not so hot.

   \begin{figure*}
   \centering
   \includegraphics[width=1.75\columnwidth]{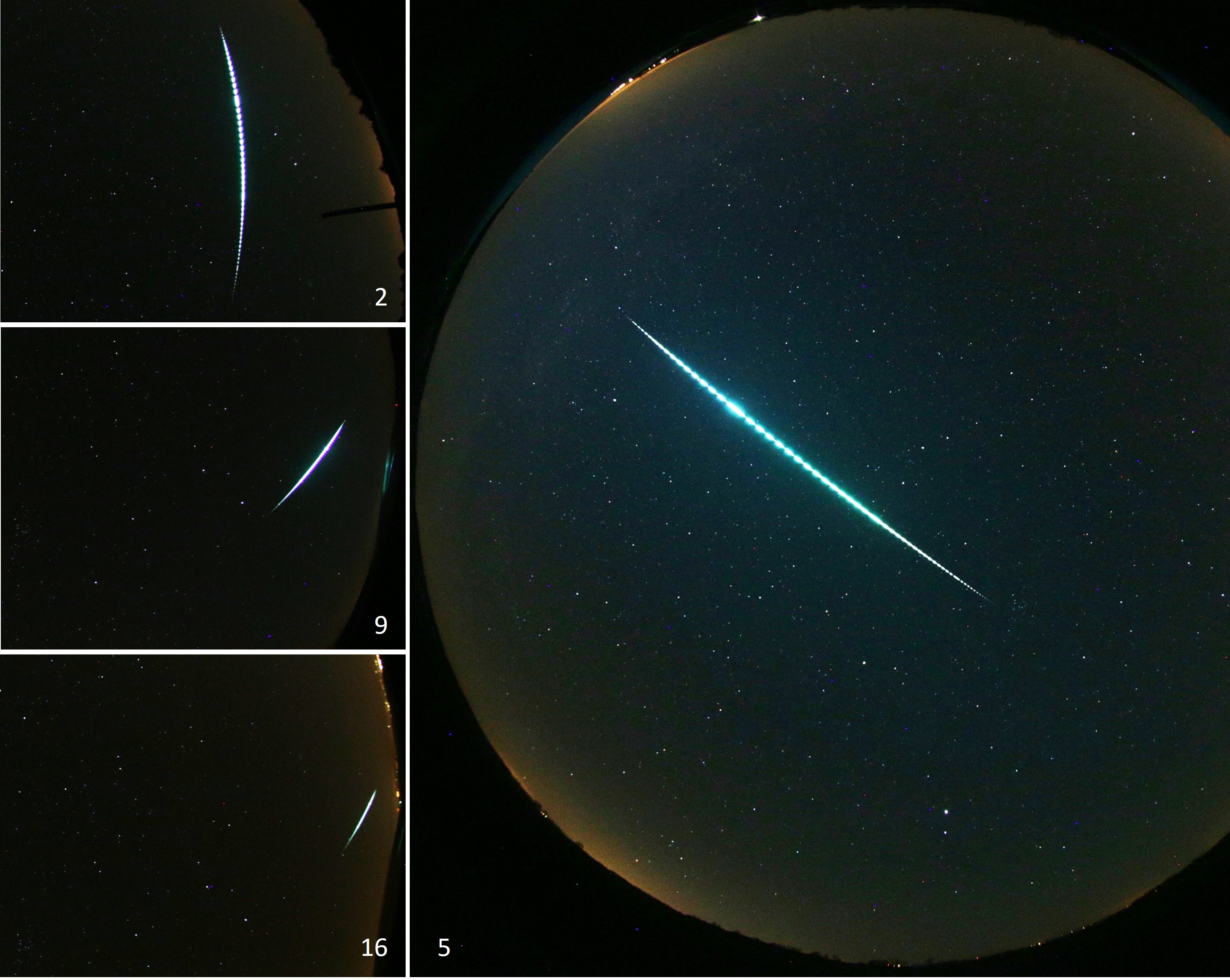}
   \caption{Images of fireball EN270217\_023122 (February 27, 2017, 02:31:22 UT) taken at four different stations. 
   In all cases, north is up,
    west is to the right, and the fireball was moving generally from the bottom to the top. The scale is the same in all images.
   At station 5, the fireball moved close to the zenith. The other three cameras imaged the fireball from 
   increasing distances of 100 km, 160 km, and 330 km at stations 2, 9, and 16, respectively 
   (measured on the ground from the middle of the fireball). The fireball entry speed was 31 km s$^{-1}$ and the maximum 
   absolute magnitude was $-12$. }
   \label{example}
   \end{figure*}
   
The air conditioning is controlled by a microcontroller 24 hours a day. The microcontroller also switches on the main computer 
before the planned exposure. Exposures are planned for the period when the Sun is more than 6\fdg5 below horizon.
Lower ISO speeds, starting from ISO 100 and increasing as the sky becomes darker, are used during 
the evening and morning twilight. During the night, ISO 3200 is used regardless of lunar phase.
Images and radiometric data are saved to the local hard disk with a capacity of 2 or 3 TB. Images are saved both
in raw format (CR2) and in jpeg format. A long winter night can provide almost 50 GB of data.

\subsection{Network architecture and fireball detection}

There were 16 fireball stations equipped with DAFO at the beginning of 2017. By the end
of 2018, the number increased to 18. Geographical coordinates (longitude, latitude, and altitude)
of those stations in the World Geodetic System 1984 (WGS84) system are given in Table~\ref{stationlist}. 
Fourteen stations
were located in Czechia, three in Slovakia (nos.\ 23--25), and one in Austria (no.\ 26).
Some of the Czech stations have been in use since the network beginning
in the 1960s. A schematic map of the stations is given in Fig.~\ref{stationmap}. The mutual distances
of neighboring stations range from 47 km (between stations 4 and 5) to 147 km (between 23 and 24).
The maximal longitudinal extent of the network is 700 km (between 11 and 24), while the
maximal latitudinal extent is 280 km (between 6 and 26). Of course, the network observed fireballs over a
larger area.
The map in Fig.~\ref{map} shows the spatial distribution of the fireballs presented in this study.
Bright fireballs (mag $< -10$) were observed more or less uniformly
over an area of $7\times10^5$ km$^2$ ($900\times800$ km). Fainter fireballs were, naturally, observed over a smaller area,
closer to the stations.

The headquarters of the network is located at the Ond\v{r}ejov Observatory, where 
data are analyzed and archived. At the time of DAFO installations, the capacity of the internet connection
was limited on some stations, so that only images of special interest were possible to transfer. 
Other images were periodically transported on external disks. Later it became possible to transfer
all images in jpeg format and all radiometric data automatically to the central server in Ond\v{r}ejov. 

Fireball detection is possible in two ways. Bright enough fireballs (brighter than an apparent magnitude of about $-5$
on moonless nights) are detected by radiometers in real time. Detections are written to data logs at
each station and transferred to the central server at the end of each night. The brightest fireballs (magnitude $\loa -10$)
are reported by email immediately. Evaluation of data logs from different stations can reveal real fireballs
as opposed to local terrestrial sources, other celestial events (such as lightning or satellite glints), or spurious 
brightenings (such as those caused by clouds passing in front of the Moon). Final confirmation is made on
images taken at the given time. Radiometers can also detect nonimaged fireballs for cloudy nights.
The approximate location and brightness of the fireball can be judged from the radiometric data, 
but no further analysis is possible.

Although radiometers can provide quick information on bright fireballs, their real-time detection algorithm is not
configured to detect fainter fireballs present in the images (to avoid an excessive number of false events). 
Moreover, radiometer sensitivity is degraded during full Moon periods. 
All jpeg images are therefore searched independently for fireballs. The search algorithm 
based on the Hough transform provides cropped images of suspicious features which are then checked by a human to
find actual fireballs. A database of all fireball images is maintained. Images containing fireballs are archived 
in both jpeg and raw formats. Other images are archived in only jpeg format to save space. 

   \begin{figure}
   \centering
   \includegraphics[width=\columnwidth]{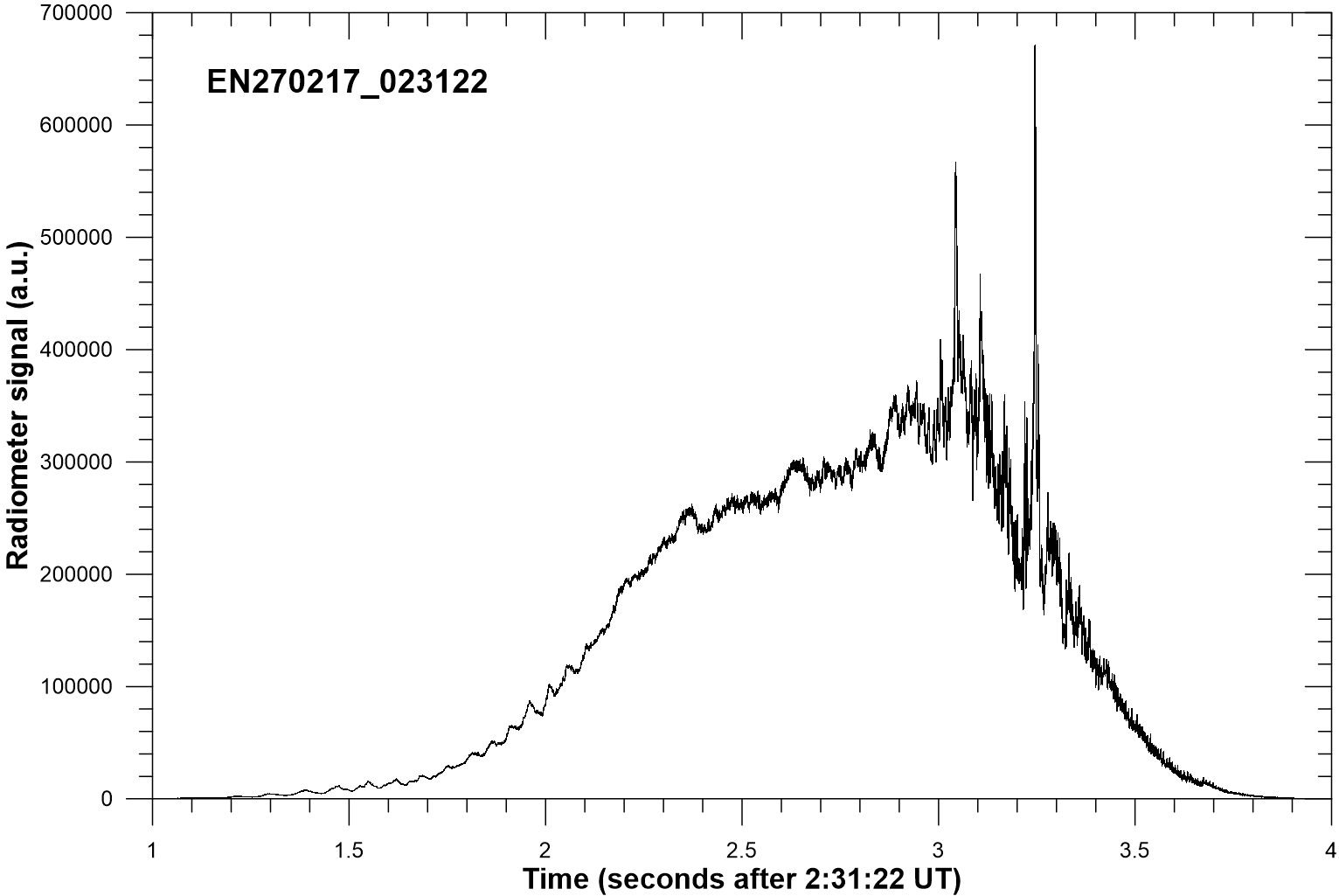}
   \caption{Uncalibrated radiometric curve of fireball EN270217\_023122 from station 5.}
   \label{radiometer}
   \end{figure}
   
Figure~\ref{example} shows, as an example, the images of fireball EN270217\_023122 taken by four
DAFOs at four different stations. Figure~\ref{radiometer} shows the radiometric curve of the same fireball.
Of course, the quality of the data depends on the distance to the fireball. The image from the most
distant station, number 16 (not used for fireball analysis), shows hardly recognizable shutter breaks. Other images show
well-defined shutter breaks including three time marks (long segments). By the comparison with the radiometric curve,
it can be easily found that the time marks correspond to times 02:31:23, 02:31:24, and 02:31:25 UT. 
The most detailed image from the closest station, number 5, shows that the breaks between the segments are empty
at the beginning and at the end of the fireball. This means that the fireball was a point-like object there. 
The parts, where the segments are not fully separated, indicate the presence of a wake or fragments
lagging behind the main body.

In addition to providing absolute timing, the radiometric curve also shows much more details 
about the fireball brightness as a function of time than the images. For example, 
there was a very short flare (full duration at half maximum $\sim$2.5 ms) on the descending branch of the light curve. 
The general shape of the light curve is not fully apparent from the image from station 5
because the fireball signal was saturated during the bright phase. It is more obvious in the images
from more distant stations, but only the radiometer can provide full details. We also note that all LCD shutters
are usually correlated across the network, so short duration flares can be missed in all images if they fall into
the shutter breaks.

\subsection{Additional cameras}

\begin{figure}
   \centering
  \includegraphics[width=\columnwidth]{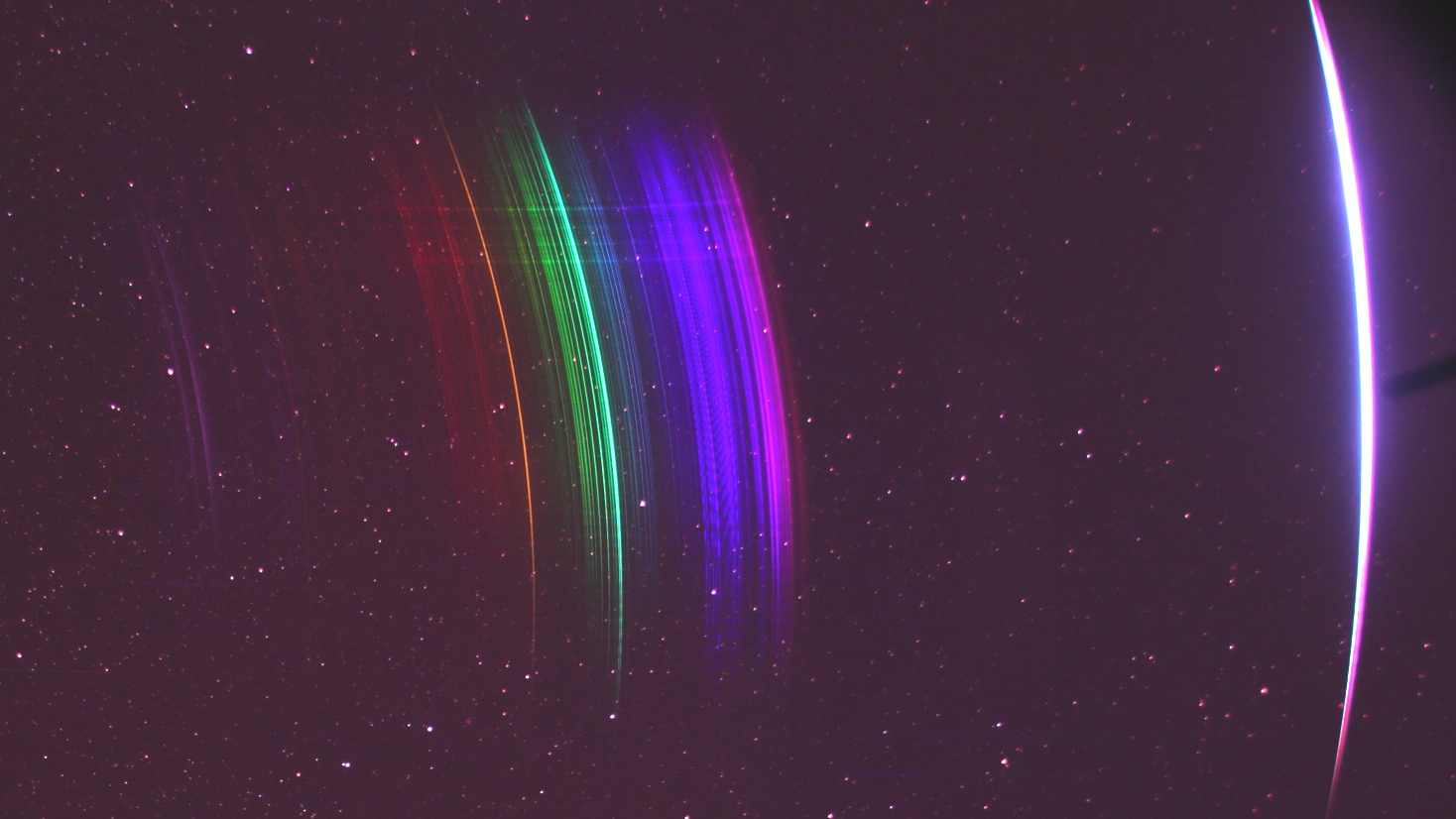}
   \caption{Fireball EN270217\_023122 (right) and its spectrum photographed by the SDAFO at station 2.}
   \label{spectrum}
   \end{figure}

Along with DAFOs, other cameras are working at some stations of the network. First, there
are spectral versions of DAFOs, called SDAFOs. They use Sigma 15mm F2.8 EX DG Diagonal Fish-eye 
lenses with holographic plastic gratings placed in front of the lenses. No LCD shutter is used.
The purpose is to capture medium-resolution spectra of bright fireballs. A more detailed description
can be found in \citet{IMC2018}. As an example, the spectrum of the fireball EN270217\_023122
 is shown in Fig.~\ref{spectrum}. SDAFOs are planned at nearly every other station. 
By the end of 2018, they were located at six stations (2, 3, 4, 10, 14, and 20).  SDAFO images
can be used to measure the fireball position using the direct (zero order) fireball image. Since there is no
velocity information and as the images have somewhat deteriorated due to the removal of the infrared filters 
from Canon cameras (to extent spectral coverage), SDAFOs have been used in the
present work rather exceptionally, for example\ when the DAFO image was not available for some reason.

Next, there are supplementary video arrays, based on Dahua IP (internet protocol) cameras, at two stations
(2 and 20). The arrays were gradually built from 2016--2018 and consist of 14 cameras at each station covering the
whole sky. The primary goal is to image fireball snapshots in flight for studies of individual fragments or wake
development. Holographic plastic gratings are attached as well, so that bright fireballs produce video spectra, 
albeit with a low dynamic range of 8 bit. There are cameras of different types, but the typical field of view
is $56\degr\times32\degr$ and the resolution is $2688\times1520$ pixels. 
More details can be found in \citet{IMC2018}.

The IP cameras were found to be useful for the measurement of velocities of some fireballs with slow angular 
motion which did not produce resolvable shutter breaks in DAFOs. There is also a narrow field video camera on  the
fast movable mount used to track fireballs in flight and image their fragmentation. This system,
called the Fireball Intelligent Positioning System (FIPS), is placed on station 2 and 20. 
It was, nevertheless, not used for the present work.

Stations 2 and 20 also host image-intensified video cameras of the Meteor Automatic Imager and Analyser (MAIA)
system which is not part of the fireball network. It is used for observations of faint meteors \citep{MAIA}. 
Because of much higher sensitivity, a fireball starting in the field of view can be captured much earlier, that is\
at higher altitudes, than by DAFO or IP cameras. Data from the MAIA system were used for a couple of fireballs
in the present work.

Finally, some casual images or videos, for example\ provided by amateur astronomers, were used as well. This concerns
important fireballs with predicted meteorite falls. These fireballs were analyzed in more detail
in \citet{2strengths}, where more details can be found.

\section{Image measurement and computation of fireball parameters}
\label{reduction}

In this section, the measurement of DAFO images, which provide the vast majority of the data, is described. 
The methods of computing fireball trajectories, orbits, light curves, and other data are also outlined.
The incorporation of radiometric data and data from additional cameras is also explained. We used two 
main software tools developed by us. Program Fishscan was used to measure photographic images and 
video records containing 
meteors. The measurements are saved in a human readable text file called MED (MEteor Data). 
Program Boltrack reads MED files and was used to compute the fireball data presented in this work.
Both programs are graphical interactive tools for personal computers using the Microsoft Windows operating
system. Another program, called Djas, was used to extract the fireball signal from raw radiometric data. 
The output is also saved in MED format and read by Boltrack.

\subsection{Image measurement}

The raw DAFO images were converted for the measurement from CR2 format into 16-bit tagged image file (TIF) format 
using Canon tools\footnote{Digital Photo Professional}. The images were cropped to a 1:1 aspect ratio
and intensities are saved in linear scale (the pixel value is proportional to incident light).
The same operations were performed with the dark image. Dark images were taken by DAFOs at the end
of each observing night or when the observation had been suspended because of bad weather.
They were taken with a closed lens cover and iris set to maximum. 

Upon reading by Fishscan, the color image was converted to grayscale using the common formula
$ S=0.30\,R + 0.59\,G + 0.11\,B$, where $R,G,B$ are pixel values in a red, green, and blue color,
respectively, and $S$ is the resulting grayscale signal. Then the dark image was subtracted and the image
was flat-fielded to account for lens vignetting. The function $S\!_{\rm f} = S/\sqrt{1-(r/12.7025)^2}$ was used for
the flat field, where $r$ is the radial distance from the image center in millimeters and  $S\!_{\rm f}$ is the
flat-fielded signal.

The following measurements were performed: astrometry, stellar photometry, meteor positions, meteor velocity,
and meteor photometry. Star positions can be measured either manually or automatically. Usually, automatic centroiding
based on the center of the gravity method was used. The operator just chooses the limiting magnitude of stars
to be measured in the selected part of the image and then removes stars which were identified incorrectly
by the automatic procedure, if there are any. Typically, several hundreds of stars were measured on the whole image. In the case
that the image with a fireball does not contain enough stars, for example\ because of cloudiness or twilight conditions,
another image from the same camera can be combined with the current image, provided that the camera
did not move in between.

Stellar photometry is usually done automatically as well, by the aperture method. The background sky signal is
measured in the vicinity of each star. The same stars as for astrometry, or part of them, were used. Stars of medium
brightness (magnitude +1 to +4) are the most useful. Bright stars are usually saturated. The measured stars should be spread
over a wide range of zenith distances to enable the computation of the extinction coefficient. Images with significant
cloudiness were used for photometry only if there was no other option (i.e.,\ a better image from another station).

Meteors were measured manually. The large variety of fireball appearances on the images, produced by their variable
light curves combined with shutter breaks, prevented us from developing a reliable and universal algorithm for
meteor measurements. Manually measured meteor positions include the beginning and the end of the meteor,
that is\ the points where it started and ceased to be visible. If the direction of meteor motion is not obvious from its
appearance, it can be determined by inspecting images from several stations and estimating the position of the radiant.
Other meteor positions were measured by the user along the whole meteor path and were then used to compute the meteor
trajectory. Well-defined parts of the meteor, where the centerline of the meteor streak can be  picked out reliably, were selected.

Independently of meteor positions, shutter breaks were measured for the purpose of computing the meteor velocity. 
If shutter breaks appear point-like (as at station 16 in Fig.~\ref{example}), their centers are measured.
If they are dash-like (as at station 24 in Fig.~\ref{example}), the leading edges are measured. Trailing edges were
not measured because they may be affected by a wake or fragments and they may therefore not represent 
the position of the main body. We note that the edge measurement must take the size of meteor image into account.
The correct position is the center of the circle inscribed to the edge. Measuring the extreme edge would lead to
an artificially enhanced velocity in cases of rapidly brightening fireballs, whose radius on the image is increasing.
If there is at least one time mark (long dash) in a meteor image, the absolute time of one shutter break
can be written into the data (provided that the absolute timing of time marks is known from the radiometer or
another source).

Finally, meteor brightness was measured, basically by aperture photometry of individual shutter breaks.
In practice, scanning the shutter break along a path perpendicular to the meteor is performed and
the resulting profile is integrated. The procedure is semi-automatic. The operator ensures that
the whole meteor signal is included, the sky background is correctly subtracted, and interfering
stars are avoided, if possible.

Fireballs are typically measured on four images from four different stations, if available. If shutter breaks are not
resolvable, only meteor positions are measured. The minimum is two stations with one containing shutter breaks.
In the case of excellent data from many stations, more than four images can be measured. 
Photometry is usually measured on two best images, but sometimes on only one and sometimes on three. 
No photometry is possible for twilight fireballs when there are almost no stars on the images.
There are two such cases in the present sample.

Images from SDAFO were measured in a similar way; however, since there is no shutter, velocity and photometry
cannot be measured. Videos from IP cameras contain only brighter stars, so data from different times are often combined
for astrometry. Photometry is not done because of a nonlinear response and low bit depth of the cameras.

\subsection{Preparation of radiometric data}

Raw radiometric data have the form of time series of brightness values and the values of the control voltage.
Control voltage compensates for slow changes of sky brightness, so that brightness values remain close to a chosen 
constant level during the whole night. This arrangement makes the real time detection of brief brightness events easier. 
If a fireball appears, the brightness value increases and the voltage starts to decrease at the prescribed rate. 
In rare cases of faint (or distant) fireballs with smooth light curves without flares, their signal can be
completely nullified by this process and their presence may be apparent, only by a dip in the voltage.
In any case, the voltage change must be used to reconstruct the fireball signal. For a given incident light,
the brightness value, $B$, is proportional to $\sim V^p$, where $V$ is the voltage and the power exponent $p$
is adjusted by the operator. The usual value for TESLA V\'UVET photomultiplier is $p=7$. The background light
level, which can be constant or depend on time linearly and is determined from the signal level before and after the fireball,
was subtracted from the resulting signal. The background may contain regular fluctuations with a frequency of nearly 100 Hz
caused by artificial light sources. In that case, the fluctuations are fitted and subtracted from the fireball signal.

\subsection{Computation procedures}

The main steps of the computation procedure are the positional reduction of all images and video records, the photometric
calibration of images with measured photometry, the computation of the fireball atmospheric trajectory and
velocity, the construction of the fireball light curve using both photographic and radiometric data. 
The fireball mass and type are then evaluated and the heliocentric orbit is computed.

\subsubsection{Positional reduction}

Positional reduction means converting plate coordinates $x,y$ into azimuths and zenith distances $A,z$.
For DAFOs and SDAFOs, the all-sky conversion formulas derived by \citet{redsky} were used. The formulas 
use a double exponential function to describe radial lens distortion. The divergence between the optical axis and 
the direction to the zenith is considered as well as the nonperpendicularity of the optical axis to the sensor.
Altogether, there are 12 parameters, called reduction constants, which need to be determined. Stellar astrometry is
used for that purpose. The catalog star positions (we use the Bright Star Catalogue) ware converted from
the standard equinox to the date of observation considering precession, nutation, and proper motion.
Their $(A,z)_{\rm catalog}$ for the middle of the exposure time and the coordinates of the station were computed
and compared to $(A,z)_{\rm computed}$ obtained from the measured $x,y$ and a first approximation
of the reduction constants. The best values of the reduction constant were obtained by an iterative procedure
minimizing the difference between the two sets. It is important to note that refraction was not explicitly considered since it
is considered to be described by the reduction constants.

All results are displayed numerically and graphically. The role of the operator is to exclude 
stars with large deviations, to select a restricted procedure if all 12 reduction constants cannot be computed simultaneously,
and to solve rare situations when there is a systematic trend in residua in a part of the image. Generally,
nevertheless, DAFO images can be reduced well over the whole sky from the zenith until close to the horizon.
The standard deviation in one coordinate ($z$ or $A\sin z$) is typically between 0.005\degr\ and 0.01\degr,
which is 0.1--0.2 pixel size.

In case of IP cameras or other images and videos covering only part of the sky, the gnomonic conversion formulas
from \citet{IMC2013}, with a polynomial lens distortion, were used. These formulas contain nine reduction constants. 
In this case, the catalog stellar coordinates must be corrected for expected refraction.

\subsubsection{Trajectory computation}
\label{trajectorycomp}

Fireball atmospheric trajectory was computed from the positional measurements of the fireball on all cameras.
The straight least squares method of \citet{mimi} was used. The method assumes that the trajectory
is a straight line in space. The trajectory was computed by an iterative process minimizing the miss distances 
of individual lines of sight from the trajectory. The computation was performed in the coordinate system 
fixed to Earth's surface. The results are the apparent coordinates (right ascension and declination) of
the fireball radiant and the geographical coordinates (longitude, latitude, and height above the geoid) of the
beginning and end point of the fireball. To compute the right ascension of the radiant, a chosen time 
during the fireball appearance was used, usually the closest whole second to the fireball beginning.

The resulting deviations from the trajectory are displayed graphically. The operator can exclude
outlying measurements from computations and can set weights to individual cameras. Usually,
weights are set proportionally to the angular length of the fireball, thus cameras having captured
the fireball from a closer distance and favorable angle are preferred. If cameras other than DAFO were used, their pixel 
scale can be taken into account. If all measurements from a camera are deviating, the positional reduction can
be checked again. In exceptional cases, the whole camera can be excluded from the computation of the trajectory
(it can still be used for the computation of the velocity).

There is less freedom if only two cameras are available. In rare cases of a low convergence angle between the 
planes drawn from two stations to the fireball, and no other station available, the straight least squares method
becomes unreliable \citep{Gural}. Nevertheless, if velocity measurements are available from both stations, the
trajectory solution can be adjusted so that the velocity data from both stations become consistent. This is an
equivalent of the multiparameter fit of \citet{Gural}, but done manually and without any assumption about the
deceleration function.

The scatter of the measurements in favorable cases (well visible fireballs at distances up to $\sim 200$ km) is
typically $\la20$ m. Of course, the precision for faint or more distant fireballs is lower. Sometimes, 
the measurement of very bright fireballs may also be difficult because of their large width on the photograph.  

The fireball beginning and end points are marked on each photograph or video. The operator selects the 
measurement that is the most appropriate for the fireball. Usually, it is the measurement from the closest station to 
the fireball beginning or end, or from the station where the observing conditions were most favorable from another reason.

We note that the assumption of a straight fireball trajectory is not strictly valid. In reality, the trajectory is bent
by Earth's gravity. The curvature is directly visible in the data of long fireballs moving nearly horizontally. 
The correction to gravity was applied before computing the fireball orbit (see Sect.~\ref{orbitcomp}).

\subsubsection{Velocity computation}
\label{velocitycomp}

Once the straight trajectory has been computed, shutter break measurements can be used to compute
the position of the fireball on the trajectory as a function of time. Each measurement was projected onto
the trajectory and the distance of the projected point from the trajectory beginning point was computed.
This quantity is called the length. So, the length as a function of time was obtained. All cameras are combined
together. If time marks are available for all cameras, the time is already in absolute units. If some
time marks are missing or if there are systematic time shifts from some reason, the time scales from
different cameras can be correlated using the length data.

Velocity is not a directly measurable quantity, but is computed by fitting the time-length data by
a smooth function.  Only for a minority of fireballs is the dependency linear and can a constant velocity 
along the whole trajectory be used. Usually, a deceleration is visible in the second half of trajectory.
We prefer the physical four-parameter fit to the data based on the integrals of differential equations of
meteoroid deceleration and ablation \citep{Pecina}. The four parameters to be found are the
preatmospheric velocity $v_\infty$ (in practice computed for the height of 150 km), the ablation 
coefficient $\sigma$, the length at the initial time, $l_0$, and the quantity $Km_\infty^{-1/3}$,
proportional to the meteoroid initial mass $m_\infty$ (here the shape-density coefficient 
$K=\Gamma A \rho_{\rm d}$, where
$\Gamma$ is the drag coefficient, $A$ is the shape coefficient, and $ \rho_{\rm d}$ is the meteoroid
density). Originally \citep[e.g.,][]{Cep93}, the velocity at the initial time, $v_0$, was used as
the fourth parameter instead of $Km_\infty^{-1/3}$, but both quantities are equivalent since the
difference between $v_\infty$ and $v_0$ depends on $Km_\infty^{-1/3}$ and $\sigma$.
The atmospheric density profile was taken either from the Committee on Space  Research International  
Reference Atmosphere 1972 \citep[CIRA72,][]{CIRA} or the Naval  Research  Laboratory  Mass Spectrometer 
Incoherent  Scatter E-00 \citep[NRLMSISE-00,][]{NRLMSISE} model. Both models gave similar results for
fireball velocities. Earth's curvature was taken into account when computing atmospheric height (and corresponding
air density) for a given length \citep[see e.g.,][]{Cep93}.

In the case of only small deceleration, the four-parameter fit may not provide a solution. In that case, the fixed
value of the ablation coefficient $\sigma=0.02$ s$^2$km$^{-2}$ was used and only three parameters were iterated. 
If no solution was obtained, the operator could choose either a parabolic or combined linear-parabolic fit, with
a constant velocity until some point and then constant deceleration. As before, outlying measurements 
can be excluded and weights can be set to individual cameras. If there is a systematic discrepancy
between measurements from individual cameras, which cannot be solved by correlating times, the
trajectory solution is checked first before excluding any camera.

The obtained time-length dependency (further also called the velocity fit) is used in further computation,
for example when relating the radiometric data to fireball heights or when computing trajectory bending by
gravity. For the computation of the fireball orbit, the initial velocity is refined taking the
physical parameters of the meteoroid  into account (see Sect.~\ref{orbitcomp}).

\subsubsection{Photometric calibration}

DAFO images where stellar photometry was done are calibrated photometrically. This procedure is
independent on any fireball measurement. Only the positional reduction must be done beforehand
to enable the computation of zenith distances of stars. 

Photometric calibration is done by relating the measured signal, $S$, of a star with its visual magnitude $V$.
The relation has the form $\log S = c -0.4 bV$, where $b$ and $c$ are constants. Since DAFO images
are in linear scale, $b=1$. The sought after calibration constant is therefore only $c$. However, $V$ is not the
catalog magnitude of the star, but its apparent magnitude affected by the atmospheric extinction: $V = V_{\rm cat} + ka$,
where $V_{\rm cat}$ is the catalog magnitude, $k$ is the extinction coefficient, and $a$ is the
airmass computed from the zenith distance. The extinction coefficient is unknown, but its typical 
value $k=0.3$ can be assumed at the beginning. With that assumption, $c$ was computed as the average value
from all stars. In the case of a nonlinear detector, both $b$ and $c$ were computed by linear regression.
After that, the extinction coefficient was computed as the slope of the linear dependency of $(\log S + 0.4 b V_{\rm cat})$
on $-0.4 b a$. The value of $c$ (and possibly $b$) was then recomputed with the improved value of $k$. 
If necessary, the procedure was repeated. At both steps, outlying stars can be excluded.

The linear dependency with $b=1$ ceases to be valid even for linear detectors in the case of a saturated
signal by a bright object (star or meteor). In DAFO images, stars with a magnitude of zero show signs of saturation
after the 35 s long exposure. For fireballs, the saturation limit is at about an apparent magnitude of $-8$,
but it depends on the angular speed of the fireball. We approximated the saturation by changing $b$ to 
0.5 above a limiting signal.  The resulting dependency of a logarithm of signal
on magnitude is called a characteristic curve, though in practice we used just a broken line.
An attempt to deal with saturation more rigorously has not been made
since the light curves of bright fireballs are taken from radiometers.

\subsubsection{Constructing the light curve}

The photographic light curve was computed from the measured signals of individual shutter breaks,
known photometric calibration of the image, known positions of shutter breaks on the sky, and
the known fireball trajectory. First, the signal was converted to magnitude using the characteristic curve.
Then the ratio between the exposure time of the shutter break (1/32 s for shutter frequency 16 Hz)
and the exposure time of stars (17.5 s for camera exposure time 35 s since the shutter is closed
half the time) was taken into account. Subsequently, the resulting fireball magnitude was corrected for extinction.
Finally, the absolute magnitude was computed by converting the magnitude to the standard fireball
distance of 100 km using the inverse square law.

The calibrated photographic light curve was used to calibrate the radiometric signal and construct 
the radiometric light curve. The raw radiometric signal, $S\!_{\rm R}$, was converted to instrumental
radiometric magnitude 
$R_{\rm I} = c_{\rm R} - 2.5 \log S\!_{\rm R}$,
where $c_{\rm R}$ is an ad hoc constant.
As the radiometric signal is known as a function of time, the fireball position for any given time must
be computed. For that, the fireball trajectory and the dependency of length on time determined in
previous steps (Sects.\ \ref{trajectorycomp} and \ref{velocitycomp}) were used.
From fireball geographical coordinates and height, its zenith distance and physical distance 
from the station were computed. Zenith distance is needed since 
the radiometer response depends on the angle between the photomultiplier
axis and the direction to the source, that is on the fireball zenith distance (the photomultiplier is directed
toward zenith). The response function has been measured in laboratory. The correction can
therefore be carried out as well as the corrections to extinction and fireball distance. There is, however, a
difference between imaging cameras and radiometers for fireballs close to the horizon. Radiometers
detect not only direct light, but also light scattered in the atmosphere. Very bright fireballs can be detected
even if they are completely below horizon. For that reason, the airmass is computed differently 
for cameras and radiometers. For details see \citet{Romanian}.

After all corrections, the radiometric curve has the correct shape, but it is still shifted in magnitudes because
an arbitrary $c_{\rm R}$ was used. This constant varies among radiometers as they have different 
sensitivities and for a given radiometer it depends on the actual value of high voltage. It was therefore
determined on an individual basis using the photographic light curve. The radiometric curve was shifted
to match the photographic curve in a region where both curves have a good signal-to-noise ratio
and the photographic signal was not saturated.

It may happen for fireballs shorter than one second that there is no time mark in the photograph.
In that case, the time in the photographic light curve must be first adjusted by comparison with the
radiometric curve. Even if the radiometric magnitudes are not calibrated, common features found in
both curves (maxima, slopes, etc.) usually enable time adjustment with a precision better
than 0.1 s. The time-length dependency must be recomputed after time is shifted and, after that,
radiometric curve must be recomputed.

The advantage of radiometric curves is a much higher time resolution and much higher dynamic range,
which enables us to measure the magnitudes of the brightest recorded fireballs without signal saturation
(if signal is saturated at the closest station, more distant stations can be used).
The quality and mutual consistency of radiometric curves was demonstrated, for example, in \citet{Zdar}
for the case of the \v{Z}\v{d}\'ar nad S\'azavou meteorite fall.  Radiometric curves are
crucial, for example,\ for detailed modeling of meteoroid atmospheric fragmentation \citep{2strengths}.
Short flares can be completely missed in photographs with shutter.
On the other hand, radiometers are less sensitive than cameras and radiometric curves became noisy 
for fainter meteors. In that case, radiometric measurements can be averaged, reducing the time resolution
(e.g.,\ to 0.01 s, which is still better than 0.0625 s from photographs), but also reducing the noise.
For the faintest observed meteors, radiometric records can be completely absent, especially during
full Moon periods. Only photographic curves can then be used for the photometric analysis.
The absolute time can be usually obtained from video records.

\subsubsection{Mass estimation and classification}
\label{masscomp}

The calibrated light curve was used to compute the initial mass of the meteoroid following the
classical assumption \citep[e.g.,][]{SSR} that the radiated energy at any time is
proportional to the loss of kinetic energy in the form of mass loss. Neglecting any
terminal mass, the meteoroid initial mass was then computed from
\begin{equation}
m_{\rm phot} = \int \frac{2}{\tau(v) v^2} \,I(t)\, {\rm d}t,
\label{photmass}
\end{equation}
where $t$ is time, $v$ is velocity, $\tau$ is the luminous efficiency, and $I=I_0 10^{-0.4 M}$ is
the radiated energy. Here $M$ is the fireball magnitude and the energy of a zero-magnitude meteor
is considered to be $I_0 = 1500$~W following \citet{SSR} for V-band magnitudes and plasma temperature 
4500~K. The luminous efficiency $\tau$ (in percent) was taken from \citet{CepKiruna}:
\begin{eqnarray}
\ln \tau &=&  0.465-10.307\, \ln v+9.781\, (\ln v)^2-3.0414\, (\ln v)^3 + \nonumber \\ 
&& + \ 0.3213\, (\ln v)^4 \label{tau1} 
\hspace{1cm} {\rm for } \ \ v < 25.372,  \\
\ln \tau  &=& -1.53 +\ln v \label{tau2}
\hspace{1.5cm} {\rm for } \ \ v \ge 25.372,  \nonumber
\end{eqnarray}
where $v$ is in km s$^{-1}$ ($\ln$ is a natural logarithm). \citet{CepKiruna} provided
luminous efficiency as a function of not only velocity, but also mass. The above
equations are for a meteoroid mass of 10 kg. The mass dependence can be used
in fireball modeling, such as in \citet{2strengths}. It was found in that work that
the mass dependence is not as pronounced as in \citet{CepKiruna}; nevertheless,
$\tau$ is likely still about two times lower for small meteoroids ($\ll$1 kg) than for large
ones ($\gg$1 kg). This means that masses of small meteoroids given in this work may be 
underestimated by a factor of two. Nevertheless, it must be said that luminous efficiency 
is generally poorly known, especially for high velocities.

The mass computed from the light curve is called the photometric mass. The
dynamic mass based on measured deceleration can be computed from the velocity fit, 
namely from the parameter $Km_\infty^{-1/3}$ (see Sect. \ref{velocitycomp}). However, the
dynamic mass can correspond to the actual initial mass only in the absence of meteoroid
fragmentation, which is rarely the case. The photometric mass is therefore a much better
approximation, despite the uncertainty as to the luminous efficiency.

As found by \citet{PE}, physical properties of large meteoroids vary enormously. 
This is demonstrated by the differences in the fireball end height for otherwise
similar entry conditions (entry speed, mass, and angle). \citet{PE} proposed a criterion
based on fireball end heights called $P_E$ and defined as
\begin{equation}
P_E = \log \rho_{e} -0.42 \log m_{\rm phot76} +1.49 \log v_\infty -1.29 \log \cos z_R,
\label{PE}
\end{equation}
where $\rho_{e}$ is the atmospheric density at the fireball end height in g cm$^{-3}$, 
$m_{\rm phot76}$ is the photometric mass in grams computed from Eq.~(\ref{photmass}) using
the luminous efficiency, $\tau_{76}$, of \citet{PE}, 
$ v_\infty$ is the entry velocity in km s$^{-1}$, and 
$z_R$ is the zenith distance of the apparent radiant.
The luminous efficiency to be used when computing $P_E$ is
\begin{eqnarray}
\log \tau_{76} &=&  -0.57 
\hspace{2.72cm} {\rm for } \ \ v < 9.3,  \\
\log \tau_{76} &=&  -3.42 + 2.92 \log v 
\hspace{1cm} {\rm for } \ \ 9.3 \leq v < 12.5, \nonumber  \\
\log \tau_{76} &=&  -1.06 + 0.77 \log v 
\hspace{1cm} {\rm for } \ \ 12.5 \leq v < 17, \nonumber  \\
\log \tau_{76} &=&  -0.32 + 0.17 \log v 
\hspace{1cm} {\rm for } \ \ 17 \leq v < 27, \nonumber  \\
\log \tau_{76} &=&  -1.51 +  \log v 
\hspace{1.67cm} {\rm for } \ \  v \geq 27, \nonumber  
\end{eqnarray}
where $v$ is in km s$^{-1}$ and $\tau_{76}$ was converted here from the units
used by \citet{PE} into percent. 
We note that $\tau_{76}$ is much smaller than the modern $\tau$ (about 4.5$\times$ at
10--15 km s$^{-1}$ and 7$\times$ above 27 km s$^{-1}$), so $m_{\rm phot76}$
is much larger than $m_{\rm phot}$.

\citet{PE} defined four fireball types based on the value of the $P_E$ criterion:

\begin{eqnarray}
{\rm I} &:& -4.60 < P_E,  \label{PEvalues} \\
{\rm II} &:& -5.25 < P_E \leq -4.60, \nonumber \\
{\rm IIIA} &:& -5.70 < P_E \leq -5.25, \nonumber \\
{\rm IIIB} &:& \hspace{1.2cm} P_E \leq -5.70. \nonumber 
\end{eqnarray}
\citet{Cep88} assigned type I to ordinary chondrites of densities of 3700 kg m$^{-3}$,
type II to carbonaceous chondrites (2000 kg m$^{-3}$), type IIIA to regular
cometary material (750 kg m$^{-3}$), and type IIIB to soft cometary material 
(270 kg m$^{-3}$) present in short-period comets such as 21P/Giacobini-Zinner
(responsible for the Draconid meteor shower). These density estimates were used
when computing the preatmospheric velocity (see Sect.~\ref{orbitcomp}).

\subsubsection{Computation of heliocentric orbit}
\label{orbitcomp}

Heliocentric orbits were computed by the slightly modified analytical method of 
\citet{Cep87}.
The input values are the fireball time, apparent radiant, geographical coordinates (longitude, latitude, and height)
of an average fireball point, the velocity at that point, the initial (preatmospheric) velocity, and the beginning height.
The average fireball point and velocity were used to compute the corrections of the radiant and initial velocity
for Earth's rotation. The next correction is for Earth's gravity, that means the computation of the zenith attraction
of the radiant and the geocentric velocity. \citet{Cep87} considered the observed part of the fireball trajectory
to be a straight line and, moreover, applied the gravity correction at the average fireball height (i.e.,\ he used
the same height for Earth's rotation correction and Earth's gravity correction)\footnote{It is important to note that in the computer 
routine, he wrote that the two heights are discriminated, but in practice, when calling the routine, both were set the same.}.
Here we considered that even the luminous part of the trajectory is curved by gravity and we applied the gravity correction
at the beginning of the trajectory, where the determined preatmospheric velocity is valid.

\begin{figure}
   \centering
   \includegraphics[width=\columnwidth]{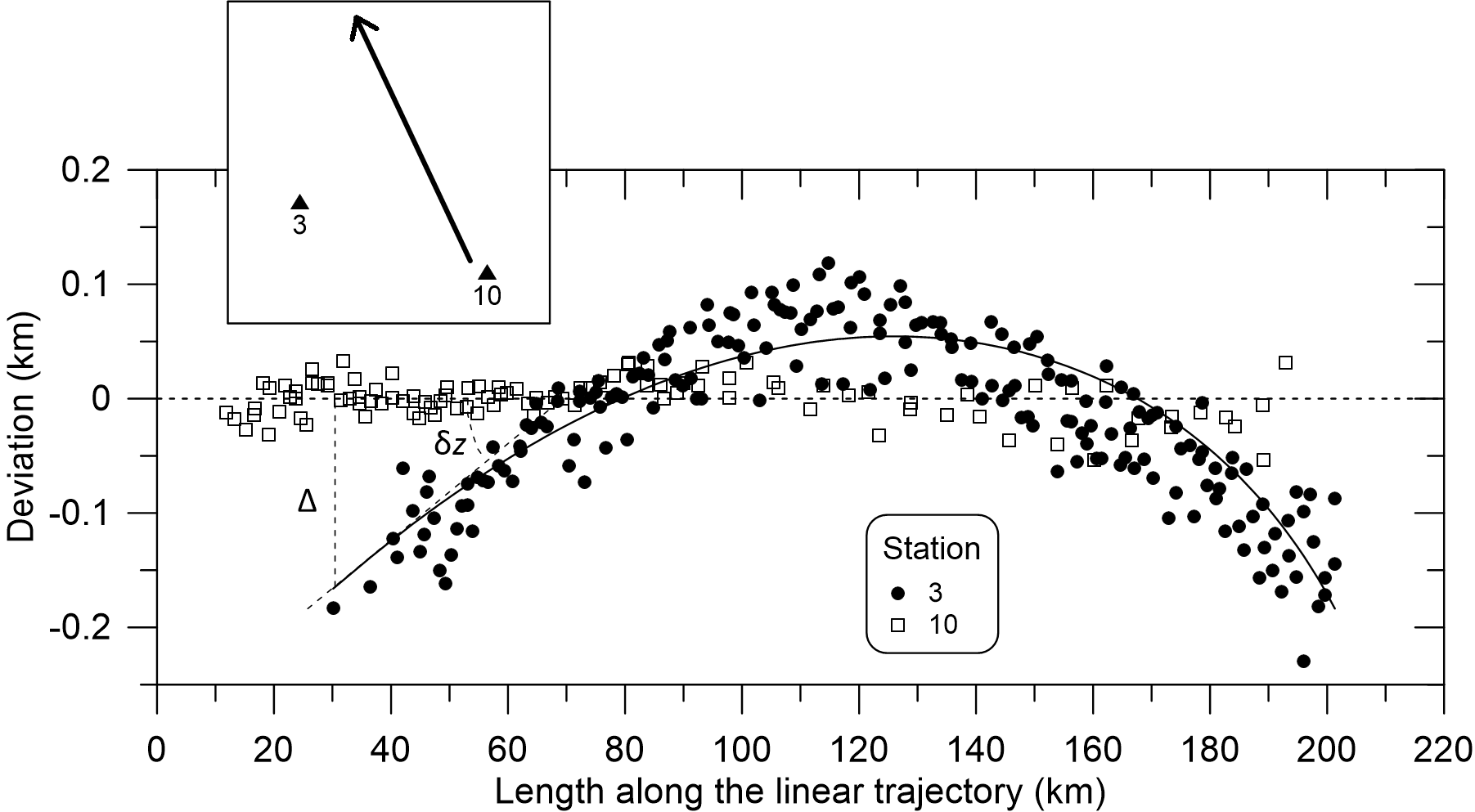}
   \caption{Deviations of lines of sight from the straight trajectory solution
   for fireball EN161216\_182218. Data from the two closest stations, numbers 3 and 10, are shown.
   The inset shows the relative positions of the stations and the fireball trajectory in a ground projection
   sketch. The fireball entered the atmosphere with a slope of 13\degr\ to the horizontal plane 
   and a velocity of 14.7 km s$^{-1}$.  The velocity was measurable for a period of 13.7~s over
   a trajectory length of 185 km.  It decreased to 8.6 km s$^{-1}$ at the end. The curved line shows
   the expected deviation $\Delta$ from the straight trajectory in the vertical plane due to gravity. 
   The angle $\delta z$ is the difference between the actual radiant, which changes along the trajectory,
   and the radiant from the straight trajectory solution.
   We note that the scale of the vertical axis is exaggerated.}
   \label{gravity}
   \end{figure}

The trajectory curvature due to gravity is a small correction to the straight trajectory solution obtained in 
Sect.~\ref{trajectorycomp}. Gravity acts in the fall plane, that is to say the\ vertical plane containing the trajectory. 
Figure~\ref{gravity} shows the observed
deviations from the straight trajectory for the long fireball EN161216\_182218 (not part of this catalog).
Station 10, which lays close to the fall plane, could not see any deviations. Station 3, which saw the fireball
from the side, detected a systematic trend which was obviously caused by gravity. The expected deviation from
the straight trajectory at a length $l$ measured in the fall plane perpendicularly to the straight trajectory is
\begin{equation}
\Delta = \Delta_0 -\frac{1}{2} g (t-t_0)^2 \sin z + (l-l_0) \sin \delta z_0,
\label{eqgravity}
\end{equation}
where $\Delta_0$ is the deviation at the beginning, that is\ at the length $l_0$ and time $t_0$; $g$ is the
gravity acceleration at the fireball height; $t$ is time; $z$ is the zenith distance of the radiant from the
straight trajectory solution; and $\delta z_0$
is the difference between the true zenith distance of the radiant at the beginning and that form the
straight trajectory solution. A negative sign means the direction to the surface. To evaluate Eq.~(\ref{eqgravity}),
the time-length dependency found in Sect.~\ref{velocitycomp} must be used. 

The two unknown quantities are $\Delta_0$ and $\delta z_0$. If the gravity bending is seen in the data,
as in case of EN161216\_182218, they are determined manually from the plot such as in Fig.~\ref{gravity}.
Only the data from stations with a favorable geometry, that is to say lying far from the impact plane, can be used.
For a majority of fireballs, however, the deviations due to gravity are smaller than the precision of the data.
In that case, $\Delta_0$ and $\delta z_0$ are computed under the assumption of symmetry. 
In particular, $\delta z_0$ is computed from the assumption that the radiant from the straight trajectory
solution is valid for the middle of the fireball. Then, $\delta z_0=g(\bar{t}-t_0) \sin z/\bar{v}$, where
$\bar{t}$ and $\bar{v}$ are the time and velocity, respectively, for the fireball middle point.

Another detail to be considered before the orbit computation is the value of the preatmospheric
velocity. Taking an average velocity in the first part of the trajectory would be inadequate
since the measured velocity can be affected by atmospheric deceleration, which can start
even before the fireball beginning, as recently stressed by \citet{Vida}. The original approach of
\citet{Cep87} was to take the $v_\infty$ value from the physical velocity fit
 (see Sect.~\ref{velocitycomp}). This value can be, on the contrary, an overestimation in the case
 of severely fragmenting meteoroids since the whole trajectory fit then gives an unrealistically small
 dynamic mass, leading to unrealistically high deceleration before the fireball start. 
 We have therefore modified the method so that only two parameters remain free, one of them
 being the initial velocity. The method was applied to the beginning of the fireball, before severe fragmentation
 can affect the dynamics. The meteoroid mass was fixed and considered to be equal to the photometric mass. 
 The meteoroid density was estimated according to the $P_E$ value (Sect.~\ref{masscomp}) or, in case
 of iron meteoroids, according to the spectrum. The ablation coefficient and $\Gamma A$ were also fixed.
 The resulting velocities have been found to be not very sensitive to their actual values 
 ($\sigma = 0.01$ s$^2$km$^{-2}$  and $\Gamma A = 1.0$ were used).

 \subsubsection{Comparison with analog cameras}
 
 The development of DAFO and the Boltrack program have been part of the continuous improvement
 of the performance of the European Fireball Network. One could ask how big of an impact these improvements
have on the final product, that is\ the fireball data,  in comparison with the situation  when AFOs were introduced
 \citep{Spurny_IAU}. We leave out obvious logistic advantages of digital cameras, such
 as the immediate availability of images for measurement (in comparison with transporting, developing, and scanning films).
 It is also clear that the quantity of the data increased dramatically (about ten times in terms of fireballs detected
 per year) thanks to the higher sensitivity of digital cameras, their ability to provide good data in poor conditions
 (such as moonlight, twilight, or partly cloudy skies), and an increased number of stations. 
 Here, we are mainly interested in the precision of the data.
 
 The basis of good data is precise astrometry. AFO used a fish-eye lens with a focal length of 30~mm, while DAFO
 uses a 8~mm fish-eye lens.  Better precision could therefore be expected for AFO. However, the practice showed that
 DAFO astrometry is at least as precise as that from AFO. One reason is that DAFO's point-like star images are easier to measure
 and more numerous than AFO's star trail beginnings or ends. In AFO images,
convenient star trails were especially hard to find close to the horizon. We often had to rely on the lens distortion
 parameters (reduction constants) determined from the guided camera \citep{redsky}. It is also possible that
films were not perfectly, mechanically stable during the long exposure or during the development process, 
worsening the astrometric performance.
 The standard deviation of the mean star position in one coordinate was typically 0.015\degr\ in AFO using $\sim$50 stars.
 For DAFO, it is about half of that value when using $\sim$200 stars.
 
 As for the meteor measurement, a higher sensitivity of DAFO means that fireballs start to be recorded earlier during their
 gradual brightening at the beginning. At the end, the difference is not so large since the drop in the brightness
 is usually steeper. In any case, a longer record is an advantage for the determination of the trajectory and the radiant, but especially 
 for the initial velocity, which is very important for the computation of the orbit. In addition, 
 velocity measurements are now combined from all cameras, improving the precision further. On the other hand, 
 very bright fireballs  (apparent magnitude $\la -13$) are more saturated and their measurement is more difficult in DAFO 
 than in AFO. Nevertheless, this effect is not severe.
 
 Photometry is more precise on DAFO thanks to the linearity of the CMOS detector (until the saturation limit). Moreover, 
 the ability of DAFO to produce time marks makes it straightforward to combine photometry with a radiometric curve
 (for fireballs longer than one second). In AFO, we had to correlate timing according to the shape of
 the light curve, which was not reliable if the meteor did not contain a natural time mark (e.g.,\ a well-defined flare).
 
 In summary, the precision of the determination of the trajectory and velocity is comparable for AFO and DAFO
 for fireballs which are bright and long enough to be well recorded in AFO and are not close to the horizon
 ($\la 15\degr$). In other cases, the trajectory and velocity precision is better from DAFO. 
 In all cases,  DAFO provides more precise photometry by combining 
 a CMOS detector with a radiometer. For an example of a bright fireball observed by both
 DAFO and AFO, see the \v{Z}\v{d}\'{a}r nad S\'{a}zavou fireball \citep{Zdar}.
 
 \section{Catalog description}
 \label{catalog}
 
The fireball catalog is provided at the CDS. There is one (long) line for each fireball.
The data contain information about the fireball trajectory, velocity, heliocentric orbit, physical
classification, and some other information, for example about observing circumstances. 
Fireball details such as full light curves or full dynamic data are not part of this catalog.
In this section, the items in the catalog are described and explained.

\textit{\textbf{Fireball code.}} Historically, fireballs within the European Fireball Network were designated EN\textit{ddmmyy},
where \textit{dd} is the day, \textit{mm} is the month, and \textit{yy} are the last two digits 
of the year of the fireball appearance. If more fireballs were within one day, they were distinguished
by additional letters (A, B, etc.). After the introduction of DAFOs, the number of detected
fireballs increased rapidly and the codes were modified to EN\textit{ddmmyy\_hhmmss},
where \textit{hhmmss} is the time (UT) of the observed fireball beginning with truncated decimal digits.
The format EN\textit{ddmmyy\_hhmm} can be used if the exact time is unknown.

\textit{\textbf{Date and time}}. The fireball appearance time is given as the year, month, day, hour, minute, and second (Universal Time, UT) 
and is valid for the average point along the trajectory.

\textit{\textbf{Time error}}. The uncertainty of the fireball time is normally given as 0.1~s. Only in exceptional
cases when there are no radiometric or video data may the uncertainty be up to $\pm15$ s. Such an uncertainty
occurs if all cameras, which recorded the fireball, started the exposure at the same time.

\textit{\textbf{Julian date}}. The fireball appearance time, valid for the average point
along the trajectory, is also given as the Julian date. 

\textit{\textbf{Solar longitude}}. The solar longitude in equinox J2000.0 at the time of the fireball is provided.

\textit{\textbf{Longitude, latitude, and height of the beginning point}}. The geographical coordinates 
of the beginning point of the fireball trajectory are given in the WGS84 
system. Trajectory bending due to gravity is taken into
account. The beginning point is selected by the operator from the beginning points measured on individual
fireball images. Usually, the image from the closest station or the station where the fireball had low angular velocity,
and thus the camera sensitivity was higher, is selected. Since the onset of brightness is slow in many fireballs,
the observed beginning height depends on circumstances such as the distance of the fireball and weather conditions. 
In two cases, when sensitive MAIA video cameras were used (EN060818\_221424 and EN140818\_223801),
the observed beginning is much higher than it would be from regular cameras. 

\textit{\textbf{Longitude, latitude, and height of the end point}}. The geographical coordinates of the end point of the fireball trajectory 
are given in the WGS84 system. Trajectory bending due to gravity is taken into
account. The end point is selected by the operator from the end points measured on individual
fireball images. Though the decrease in brightness at the end is often steeper than the increase at the beginning,
the end height is also affected by observing circumstances.

\textit{\textbf{Longitude, latitude, and height of the average point}}. In addition to the beginning and end point,
the geographical coordinates of the average point of the fireball trajectory are also provided in the WGS84 
system. Trajectory bending due to gravity is taken into account. 
The average point is near, but not necessarily exactly in, the middle of the trajectory; 
it depends on the distribution of the trajectory measurements. Meteor time is given for the average point. 
The correction for Earth's rotation, when computing the heliocentric orbit, is performed at the average point.

\textit{\textbf{Longitude, latitude, and height of the point of maximum brightness}}. Finally, the geographical coordinates 
of the point, where the fireball reached maximum brightness, is given in the WGS84 system. We note that the point of
maximum brightness is not always distinct. There may be two or more flares with an almost identical
brightness or the top of the light curve can be flat with no clear maximum. Data are not available for fireballs
without photometry (observed in early dusk or late dawn).

\textit{\textbf{Length}}. The observed length of the fireball in kilometers is measured from the beginning to the end point. 
It is taken from the linear trajectory solution since the length measured along the curved trajectory would be almost the same.

\textit{\textbf{Duration}}. The measured duration of the fireball in seconds is computed as the time difference between the
last and the first measured velocity point (i.e.,\ shutter break or video frame). The real duration was longer
in all cases. The given duration expresses the interval available for velocity measurements.

\textit{\textbf{Average azimuth and zenith distance of the radiant}}. The average horizontal coordinates of the radiant are computed as 
the arithmetic mean between radiant azimuth 
and zenith distance at the beginning and the end of the trajectory. The azimuth changes along the trajectory
because of Earth's curvature. The zenith distance changes because of both Earth's curvature and trajectory curvature.
The zenith distance also expresses the angle between the direction of motion and the vertical direction (to nadir).
The azimuth is counted from from the north to the east.

\textit{\textbf{Right ascension and declination of the apparent radiant}}. The equatorial coordinates of the fireball 
apparent radiant are given in equinox J2000.0 and are valid for the beginning point. Both right ascension and declination change
along the trajectory because of the trajectory curvature. The given errors are computed from the
spread of positional measurements.

\textit{\textbf{Initial no-atmosphere velocity ($v_\infty$)}}. The initial velocity is given with atmospheric influence subtracted and is
valid for the fireball beginning point. The meteoroid would have this velocity at that point if it is only influenced
by gravity, but not atmospheric drag. The given error takes the uncertainty of the meteoroid mass into account.

\textit{\textbf{Velocity at the point of maximum brightness}}. The velocity at maximum brightness is obtained 
from the velocity fit for the whole trajectory.
If the maximum brightness point lies outside the part of the fireball covered by velocity measurements (which happens
e.g.,\ for fireballs with a bright terminal flare), the velocity is obtained by extrapolation.
It is not available for fireballs without photometry.

\textit{\textbf{Terminal velocity}}. The velocity at the last measured velocity point is obtained from the velocity fit for
the whole trajectory.

\textit{\textbf{Height of terminal velocity measurement}}. The height of the last measured velocity point is provided. 
Depending on the observing conditions, velocity measurements may terminate some distance before the fireball
end point.

\textit{\textbf{Right ascension and declination of the geocentric radiant}}. Equatorial coordinates of the fireball 
geocentric radiant (i.e.,\ apparent radiant corrected for Earth's rotation and gravity) are given in equinox J2000.0.

\textit{\textbf{Sun-related ecliptical coordinates of the geocentric radiant}}. Ecliptical longitude minus solar
longitude and ecliptical latitude of the geocentric radiant can be used to study the so-called sporadic
meteoroid sources or the dispersion of radiants of meteor showers.

\textit{\textbf{Geocentric velocity}}. Geocentric velocity is the no-atmosphere velocity relative to Earth's center corrected
for Earth's gravity.
 
\textit{\textbf{Heliocentric radiant}}. Heliocentric radiant  is the geocentric radiant corrected for Earth's orbital motion.  
Ecliptical longitude and latitude are given.

\textit{\textbf{Heliocentric velocity}}. Heliocentric velocity is the geocentric velocity corrected for Earth's orbital motion. 

\textit{\textbf{Orbital elements}}. Elements of the heliocentric orbit  are given in equinox J2000.0. 
Although just
six elements are needed to define the orbit, ten elements are provided for convenience. They are the following: 
the {semimajor axis, eccentricity, perihelion distance, aphelion distance, inclination,
argument of perihelion, longitude of the ascending node, longitude of the perihelion, date of the last perihelion
passage, and orbital period}. We note that in the case of orbits with eccentricities close to one, the semimajor axis,
aphelion distance, and period are not well restricted and their errors may be larger than their values. 
The same is true for the date of the last perihelion passage if the Earth encounter occurred before perihelion.
In the case of nominally hyperbolic orbits, the aphelion distance, orbital period, and date of perihelion passage
are not given (are set to zero). The semimajor axis is negative for hyperbolic orbits.

\textit{\textbf{Tisserand parameter}}. The Tisserand parameter relative to Jupiter can be used for orbit classification
\citep{Tisserand}.

\textit{\textbf{Maximum brightness}}. The absolute (100 km distance) visual magnitude reached by the fireball
at the maximum brightness point is provided as the first physical parameter. 
The typical uncertainty is $\sim$\,0.1 mag. For the faintest fireballs in the set,
the uncertainty may reach 0.5 mag. We note that the maximum brightness does not always express the significance
of the fireball. Sometimes, the maximum is reached in a narrow large amplitude flare with a duration of less than
0.1 s. It is not available for fireballs without photometry.

\textit{\textbf{Total radiated energy}}. The total radiated energy is the energy radiated by the fireball to all directions 
at all wavelengths. It is obtained by the integration of the light curve using the conversion factor of 1500 W for a zero magnitude fireball
\citep{SSR}. It is not available for fireballs without photometry.

\textit{\textbf{Photometric mass}}. Initial mass of the meteoroid is computed from the light curve using
Eq.~(\ref{photmass}). It is not available for fireballs without photometry.

\textit{\textbf{Terminal dynamic mass}}. An estimate of the meteoroid mass at the last measured velocity point is
computed from the physical four-parameter velocity fit for the whole trajectory, under the assumption of
$\Gamma A = 0.7$ and a meteoroid density of $\rho_{\rm d} = 3000$ kg m$^{-3}$. It is given only if the
estimate is $>1$ gram and if the terminal velocity is $<10$ km s$^{-1}$, otherwise it is set to zero. 
At larger velocities, the meteoroid can be expected to be ablated out completely. 
The given mass can be comparable to the expected meteorite mass only if
the terminal velocity is about 5 km s$^{-1}$ or less. In any case, the whole trajectory fit is only approximate.
A better estimate of the mass of a possible meteorite can be obtained by meteoroid fragmentation modeling using not
only dynamics, but also a light curve \citep{2strengths} and that estimate is then used for meteorite searches.  

\textit{\textbf{$P_E$ value}}. The value of the PE criterion defined by \citet{PE},
see Eq.~(\ref{PE}), is used for the physical classification of meteoroids. It is not available for fireballs without photometry.

\textit{\textbf{Type}}. Fireball type is selected by the operator primarily according to the $P_E$ criterion.
If the $P_E$ value is near the boundary, a combined classification can be given, for example\ II/IIIA. 
It is not available for fireballs without photometry.

\textit{\textbf{Maximum dynamic pressure}}. 
Dynamic pressure is defined here as $p=\rho v^2$, where $\rho$ is the density of the atmosphere
and $v$ is the fireball velocity. The whole trajectory is considered. The velocity fit is extrapolated
to the parts of the trajectory not covered by velocity measurements.

\textit{\textbf{Height of maximum pressure}}. The height at which the maximum dynamic pressure was reached
is provided.
In the case of no deceleration, the maximum pressure is reached at the end of the trajectory. Otherwise,
it can be earlier. 

\textit{\textbf{Pf value}}. The value of a newly proposed criterion, called the pressure factor, 
is an alternative means for the physical classification of meteoroids. See Paper II. It is not available for fireballs without photometry.

\textit{\textbf{Pf-class}}. Meteoroid class is computed according to the classification based on the $P\!f$ value. See Paper II for the
definition of classes.
It is not available for fireballs without photometry.  

\textit{\textbf{Possible meteor shower}}. Three letter codes of meteor showers assigned by the International Astronomical Union (IAU) 
are used to identify the shower to which the fireball may belong. The parameters of meteor showers were
taken from the IAU Meteor Data Center (MDC) webpage\footnote{https://www.ta3.sk/IAUC22DB/MDC2007/} in 2016.
Both established showers, and showers on the working list with IAU numbers up to 821 were considered. 
While the shower membership listed here is obvious for
well-defined major showers, it is only tentative for minor showers. It was selected by the operator on the basis of the
similarity of the radiant and the orbit. It is important to note, however, that there are large differences between the published orbits
for some showers. Moreover, some showers from the working list may in fact not exist at all.

\textit{\textbf{Possible related body}}. Name of asteroid or comet with a similar orbit is listed for some fireballs. 
In cases of major showers, the listed body is the well-known parent body of the shower. 
In other cases, the association is only tentative and selected by the operator.

\textit{\textbf{Number of cameras}}. To evaluate observation circumstances, the number of different cameras 
used to compute the fireball trajectory is given.
In most cases, the number of cameras corresponds to the number of stations. 
Only in a few cases was more than one camera used at one station (e.g.,\ both the DAFO and IP camera) 
and was the number of stations thus smaller than the number 
of cameras. Of course, the minimum is two stations.

\textit{\textbf{Minimal distance}}. Another benchmark of observation circumstances is the minimal distance 
between the fireball and one of the cameras used to compute its 
trajectory. The distance was computed in three dimensions.

\textit{\textbf{Spectrum}}. The existence of the fireball spectral record, either by the SDAFO or by IP cameras,
is noted in the last column.
If ``poor" is given, only a faint spectrum or a spectrum of inferior quality with maximally three spectral lines is available.
The spectra indicating anomalous meteoroid composition are marked either ``Fe" (only lines of iron are present), 
``Na+" (the sodium line is brighter than expected), or ``Na$-$" (the sodium line is faint or missing), but they are only a few.
Most spectra are broadly consistent with a chondritic composition \citep[see][for a description of normal
fireball spectra]{2components}. A more detailed analysis of good spectra with the aim of revealing minor
chemical differences is planed for the future. The fireballs showing only Fe lines were studied by \citet{irons}.

   \begin{figure}
   \centering
   \includegraphics[width=1.0\columnwidth]{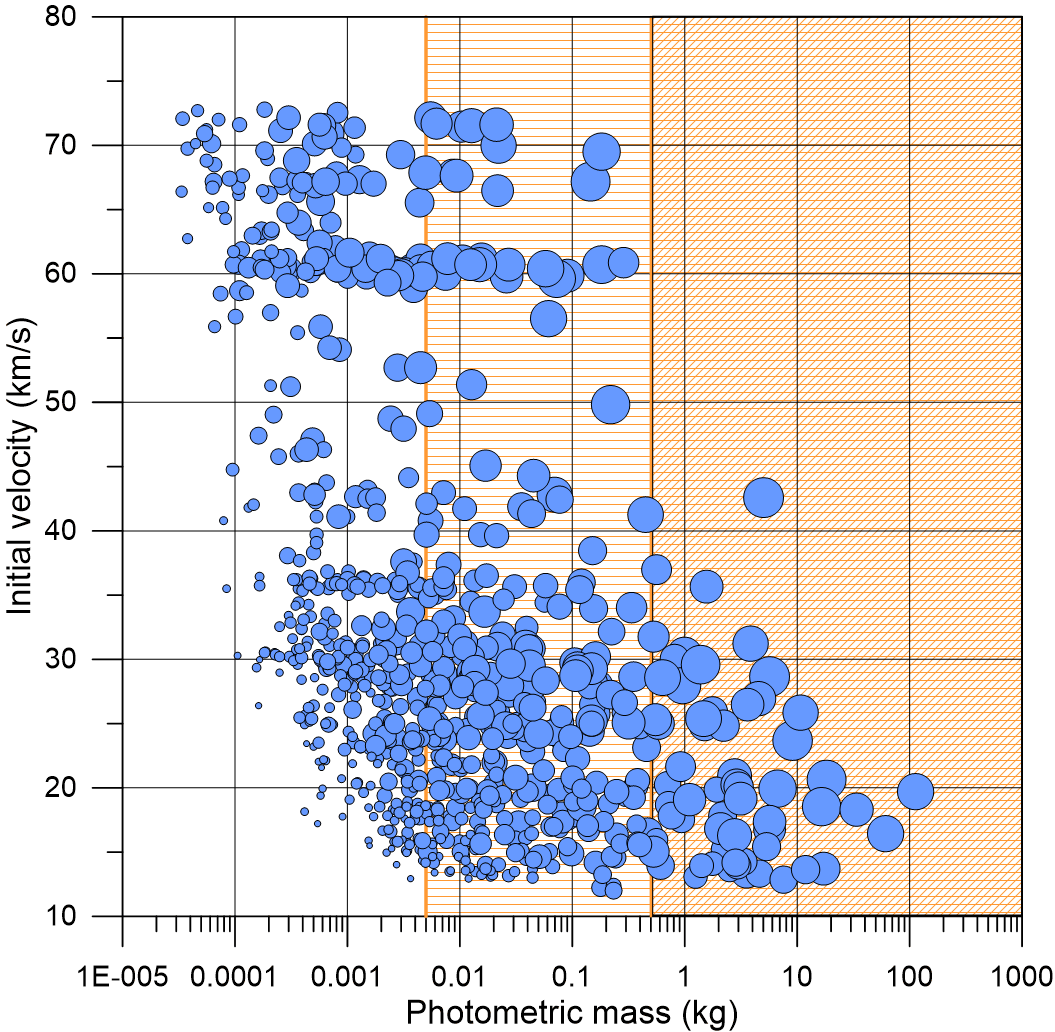}
   \caption{Relation between the meteoroid mass, velocity, and fireball brightness. The symbol size is proportional to the
   maximum brightness. The meteoroid subsamples with masses $> 5$ g and $> 0.5$ kg are marked by hatching.
   The uncertainty of velocities is lower than 1 km s$^{-1}$. The uncertainty of masses is driven 
   by the uncertainty of the luminous efficiency and may be about 30\%. Moreover, the mass of small 
   meteoroids may be underestimated by a factor of two, as explained in Sect.~\ref{masscomp}.}
   \label{masses}
   \end{figure}

\section{Description of the sample}
\label{description}

The EN database contains 2566 meteors observed by the digital cameras from at least two stations in the years
2017--2018. Most of them, however, are too short or too faint to obtain precise trajectories and orbits. They can
be used for statistical purposes only. Trajectory computations were done for 872 fireballs. Some of them were then excluded
since the results were not reliable enough. In particular, all fireballs with a minimal distance larger than about 350 km were 
excluded. Finally, 824 fireballs were included in this work and catalog.
In Appendix~\ref{histograms}, histograms of some observable quantities are presented. 
We note that some of the fireballs were already
used, together with fireballs from other years, in specialized studies, namely about meteoroid fragmentation 
\citep{2strengths, Shrbeny_wake}, iron meteoroids \citep{irons}, as well as Taurid \citep{Spurny_Bratislava, Taurid_phys}, 
September $\epsilon$ Perseid \citep{Shrbeny_SPE}, and $\eta$ Virginid \citep{Brcek} meteor showers.

Fireball brightness primarily depends on meteoroid mass and velocity. These three quantities
are compared in Fig.~\ref{masses}. It can be seen that small meteoroids were detectable only if they had a high
velocity. Only meteoroids more massive than about 5 grams (an equivalent diameter of 1.5 cm for a density of 3000 kg m$^{-3}$
or 2.5 cm for 600 kg m$^{-3}$) were observable at any velocity. It must be, 
nevertheless, kept in mind that small and slow meteoroids among them were observable over smaller area than 
the larger or faster ones. Meteoroids more massive than 0.5 kg (an equivalent diameter of 7 cm for a density of 3000 kg m$^{-3}$
or 12 cm for 600 kg m$^{-3}$) are hereafter called "large." All of them were observable over the whole covered area.
 Our sample contains 388 meteoroids larger than 5 g, and 63 of them are larger than 0.5 kg.
The largest one had a mass of about 110 kg and a probable diameter of about 40 cm. The smallest observed, fast meteoroids
with masses around $5\times10^{-5}$ kg had diameters of about 5 mm.
We note that no large meteoroid faster than 45 km s$^{-1}$ was observed and only a few faster than 30 km s$^{-1}$ were observed 
during the two years covered by this work.

   \begin{figure*}
   \centering
   \includegraphics[width=1.85\columnwidth]{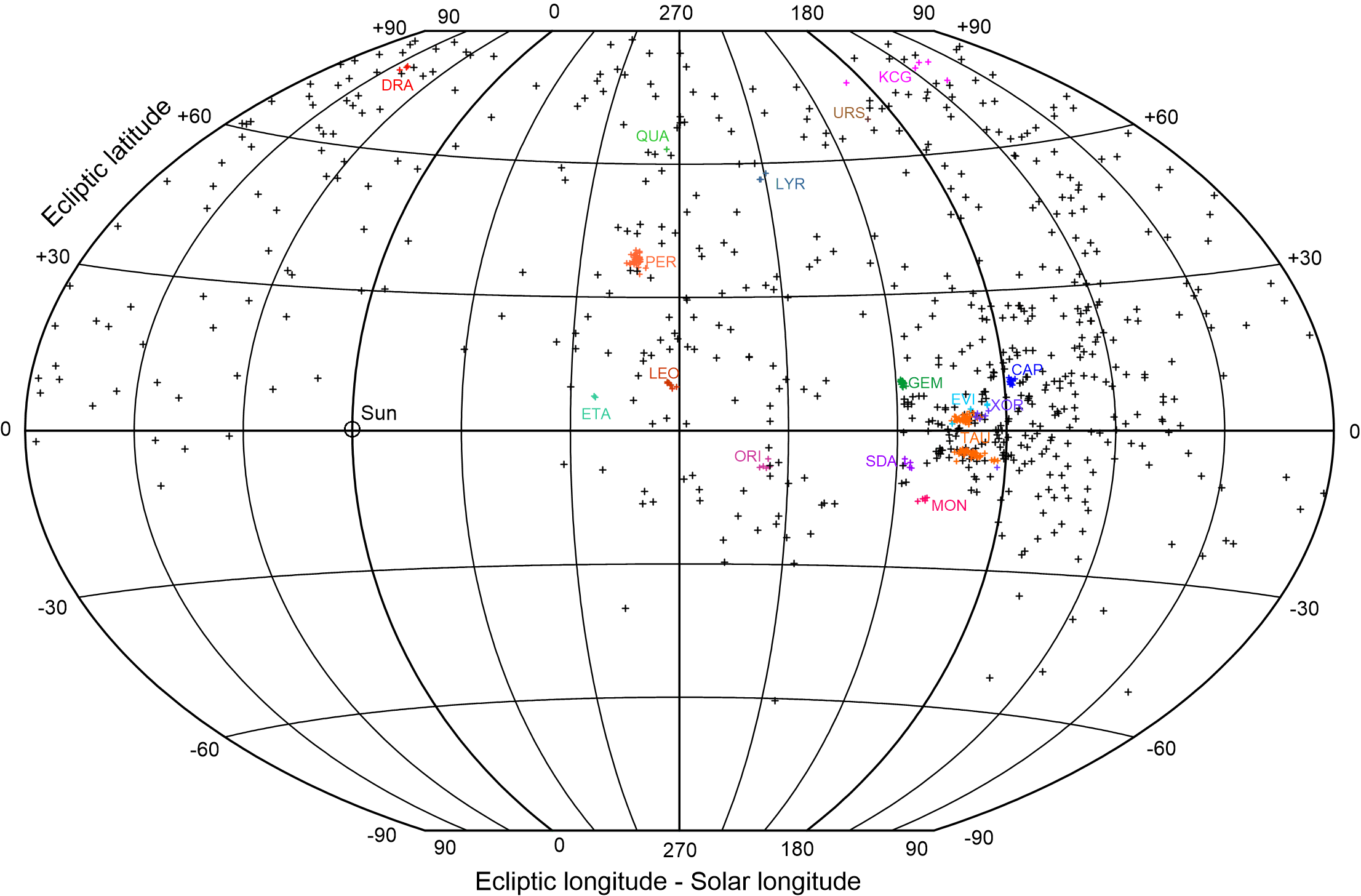}
   \caption{Geocentric radiants of all fireballs in ecliptic coordinates. The eclitptical longitude is given relative to the solar longitude at
   the fireball appearance. Radiants of fireballs assigned to 16 major meteor showers are plotted in color. The whole sky is plotted in the Winkel triple projection.}
   \label{radiants}
   \end{figure*}
   
The temporal distribution of fireballs over the year was uneven. There are peaks associated with meteor
showers Taurids in November, Perseids in August, and Geminids in December.  
The showers, nevertheless, disappear when only large meteoroids are considered
\citep[the situation may be different in the years with enhanced Taurid activity, see][]{Spurny_Taurids}.
The number of fireballs is further influenced by observational factors such
as the length of the night (substantially different in June and December at 50\degr N latitude) and weather,
which is usually better from June to August, but it varies from year to year. 
For example, the network recorded 136 fireballs in January 2017 (48 of them entered this catalog), 
but only 29 in January 2018 (11 in the catalog; in both years the weather was poor for Quadrantid maximum).
Over shorter periods of time, the lunar phase can influence the number of fainter fireballs.

In the whole sample, 222 fireballs belonged to one of 16 major meteor showers. Namely, there are 80 Taurids,
46 Perseids, 38 Geminids, ten $\alpha$ Capricornids,  eight Leonids, six $\chi$ Orionids (five northern and one southern one), 
five Orionids, five December Monocerotids, five Southern $\delta$ Aquariids, five 
$\kappa$ Cygnids, four $\eta$ Vriginids, three Lyrids, three October Draconids, two $\eta$ Aquariids, and one
Quadrantid and Ursid. We counted fireballs classified as Southern and Northern Taurids and 
also $\lambda$ Taurids (LTA), $\omega$ Taurids (FTA), and $\tau$ Taurids (TAT) into Taurids.
Roughly speaking, 73\% of fireballs observed from 2017--2018 were sporadic or belonged to minor
showers; 10\% were Taurids, 5\% Perseids, 5\% Geminids, and 7\% include all of the other major showers together. It is worth noting that,
as reported by \citet{Spurny_Bratislava}, the new branch of Taurids discovered in 2015 \citep{Spurny_Taurids} 
was also present in 2018, though at a lower activity level.

Figure~\ref{radiants} shows the distribution of radiants in ecliptic coordinates relative to the position of the Sun.
Sporadic radiants are distributed relatively evenly over the observable part of the sky (which excludes the far southern sky and is
biased against the daytime sky). Sporadic meteoroid sources are much less pronounced than in the radar data covering much smaller
meteoroids \citep{CMOR}. Only the antihelion source (180\degr\ from the Sun) shows a somewhat enhanced
concentration of radiants. Most shower radiants seem to be tightly concentrated in these kinds of coordinates.
Taurids and $\eta$ Vriginids have elongated radiants and only $\kappa$ Cygnids have a dispersed radiant. It is important to note,
however, that shower assignment may be uncertain for showers that are not defined very well, such as
$\eta$ Vriginids  and $\kappa$ Cygnids.

\section{Conclusions}

This paper has presented a catalog of 824 fireballs observed by the European Fireball Network. 
The catalog is available in electronic form at the CDS and 
the data belong to the most reliable fireball data currently available.
They can be used for various studies of meteoroids in the Solar System. This paper contains an explanation of
how the data were collected, the methods of their analysis,
a detailed description of catalog entries, and a basic statistical characterization of the dataset.
Selected scientific analyses are presented in the accompanying Paper II.

The network continues the observations. The number of cameras has increased since 2018.
The measurement and processing of the most important fireballs continues on a daily basis. More data
will be made available in the future.

\begin{acknowledgements}
We thank all local staff at the camera stations for their supervision of camera operation 
and their assistance in case of problems; namely (in alphabetical order) P.~Adam, P.~Bado\v{s}ek, 
P.~Ba\v{r}inka, Jan Bedna\v{r}\'{i}k, Ji\v{r}\'{i} Bedna\v{r}\'{i}k, M.~\v{C}erm\'ak, L.~Dobi\'a\v{s}, M.~Drechsler, M.~Hus\'arik,
L.~Kaz\'{i}k, K.~Ko\v{s}\'{\i}k, R.~Kova\v{r}\'{i}k, R.~K\v{r}enek, M.~Kudrna, I.~Kudzej, V.~Luk\'a\v{c}, J.~Macura, A.~Nov\'ak,
M.~Gregor, S.~Ondruch, L.~Paseka, J.~Pokorn\'y, M.~P\v{r}edota, F.~Putala, P.~Rapav\'y,
J.~S\'adovsk\'y, J.~S\'adovsk\'y, Jr., V.~Sv\v{e}tl\'{\i}k, R.~Szpuk, O.~\v{S}lof\'ar, D.~Tomko, and Z.~Vondra.
Students E.~\v{R}ezn\'{\i}\v{c}kov\'a, A.~Br\v{c}ek, and R.~Marhold helped us with computations.
We are obliged to the company Space Devices, s.r.o., for the development and productions of DAFOs
according to our needs and specifications. P. Koten provided data from MAIA cameras for two fireballs. 
This work was supported by grant no.\ 19-26232X from Czech Science Foundation and
Praemium Academiae of the Czech Academy of Sciences, which provided funds for digitization of the main part
of the European Fireball Network. 
The institutional research plan is RVO:67985815.
This work has also been supported, in part, by the VEGA -- the Slovak Grant
Agency for Science, grant No. 2/0059/22.
\end{acknowledgements}

\begin{appendix}

\section{Histograms of selected quantities}
\label{histograms}

In this appendix we present histograms showing the distribution of selected quantities such as the fireball velocity, 
maximum magnitude, entry angle, and orbital elements. The histograms simply show what the properties
are of meteors that are observable with our cameras, that is\ to say brighter than a magnitude of about $-2$. Not only the whole sample,
but also two subsamples with meteoroid photometric masses $>\,$5 g and $>\,$0.5 kg are presented. 
Additionally, sporadic meteoroids, that is\ those that do not belong to any of the 16 major meteor showers 
listed in Sect.~\ref{description}, are marked, irrespective of mass.
\vspace{2.8cm}

\begin{figure}[h]
\includegraphics[width=\columnwidth]{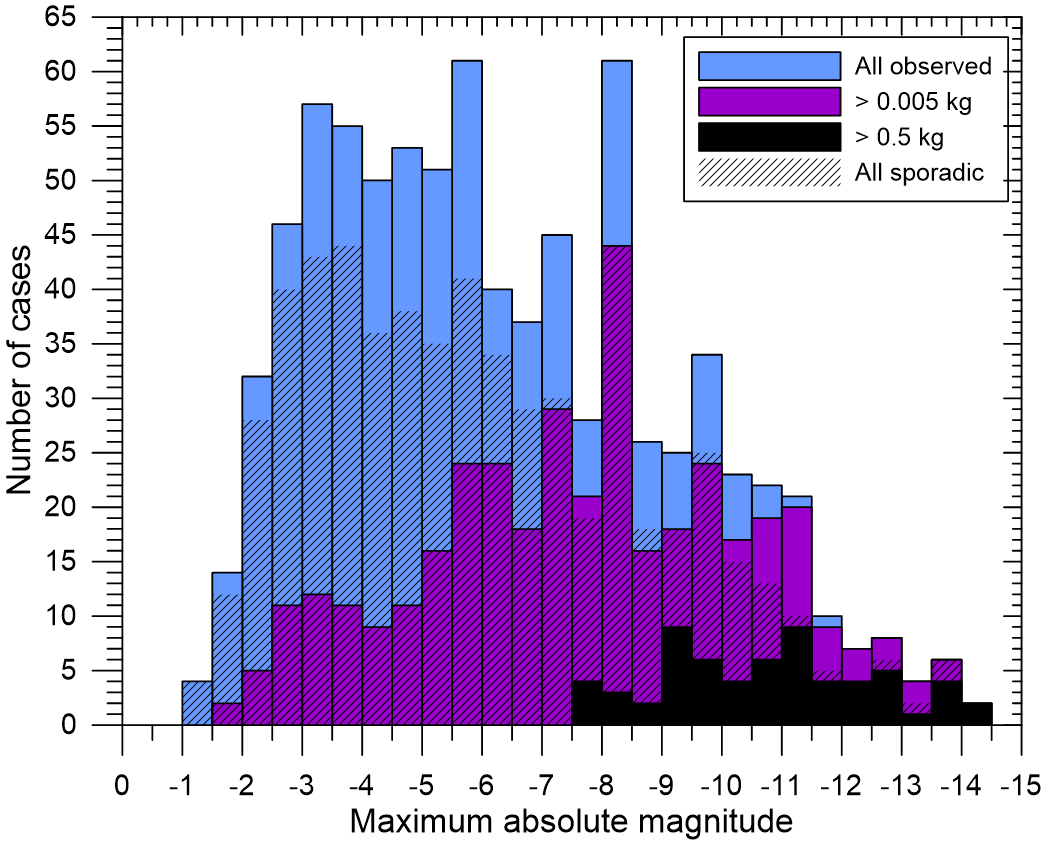}
\caption{Distribution of observed fireball magnitudes.}
\label{hist_mag}
\end{figure}

\begin{figure}[h]
\includegraphics[width=\columnwidth]{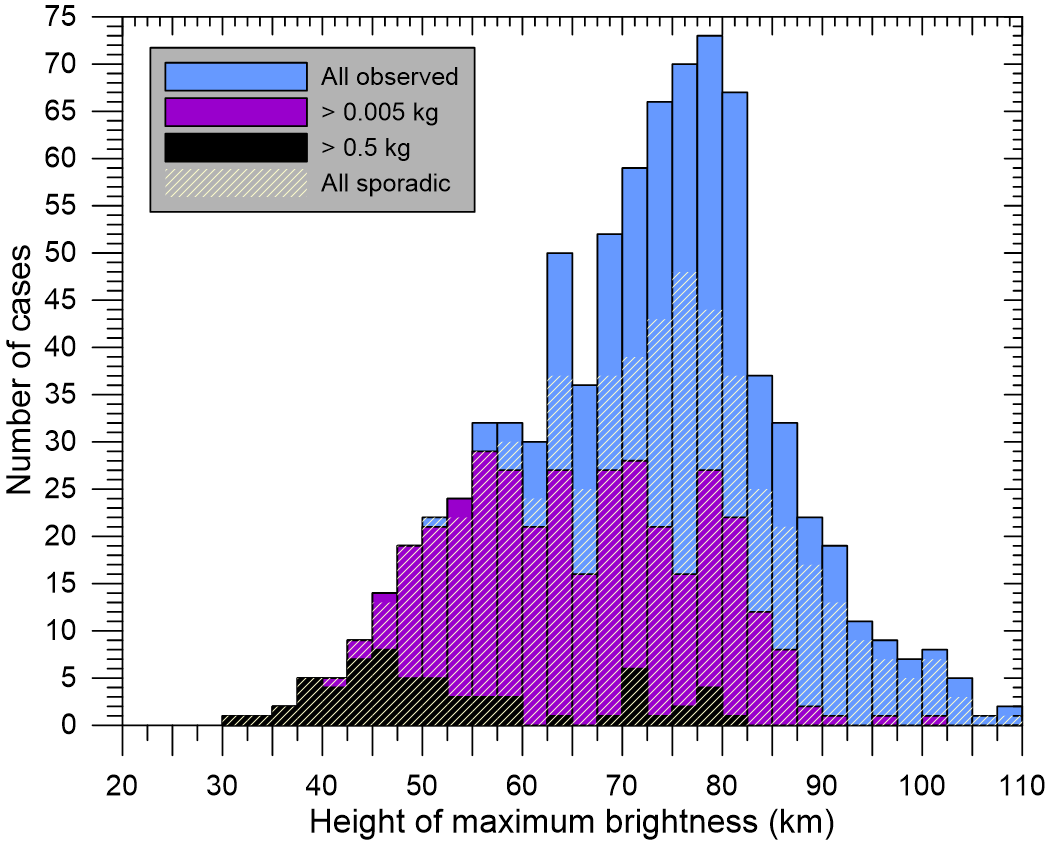}
\caption{Distribution of heights of fireball maximums.}
\label{hist_hmax}
\end{figure}

\begin{figure}
\includegraphics[width=\columnwidth]{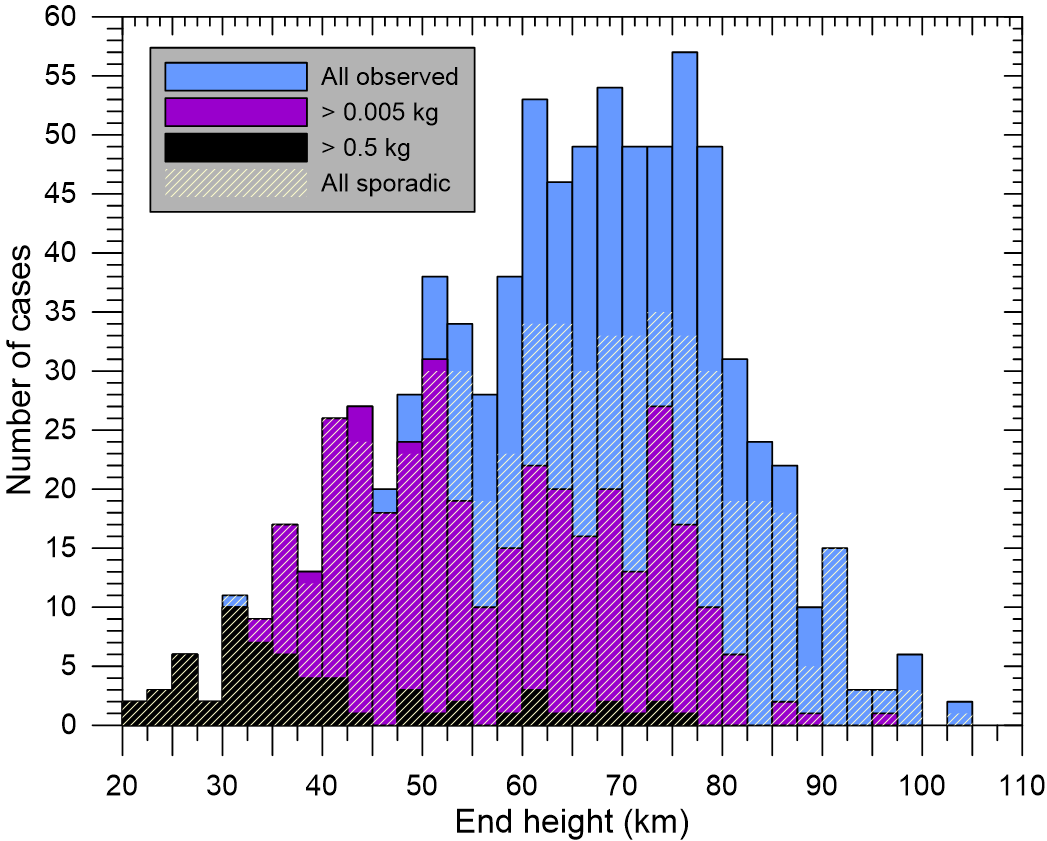}
\caption{Distribution of fireball end heights.}
\label{hist_hend}
\end{figure}

\begin{figure}
\includegraphics[width=\columnwidth]{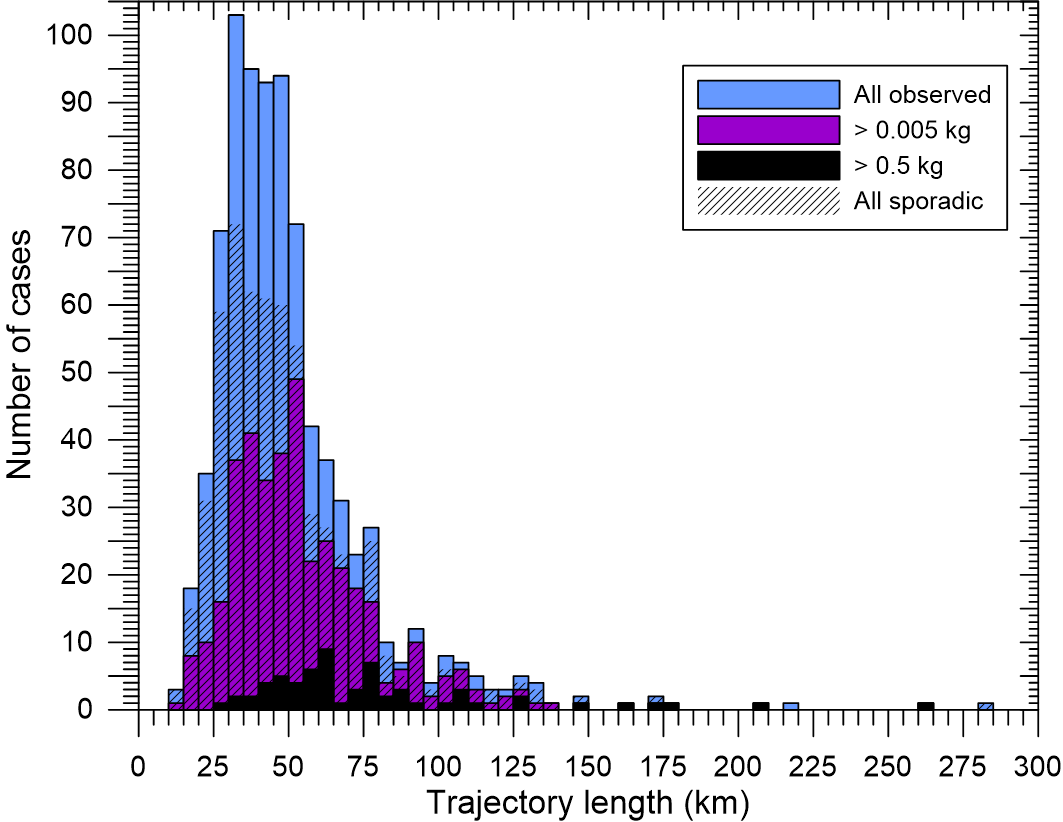}
\caption{Distribution of trajectory lengths.}
\label{hist__length}
\end{figure}

\begin{figure}
\includegraphics[width=\columnwidth]{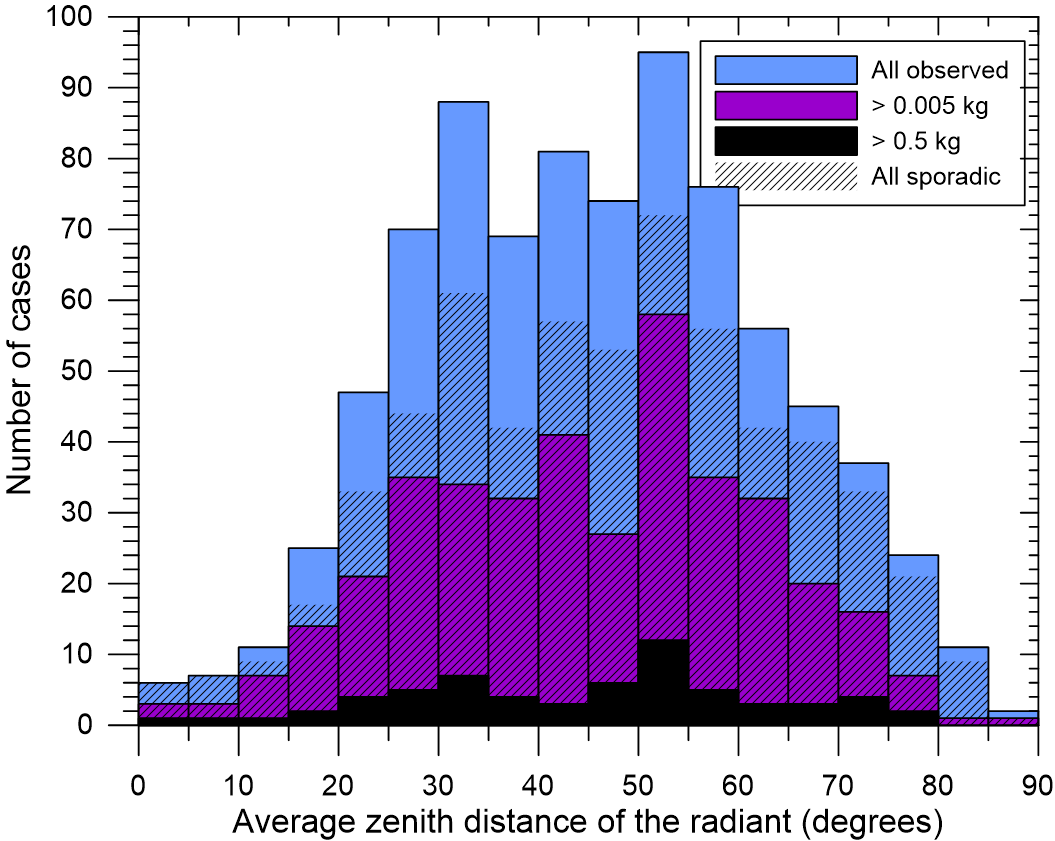}
\caption{Distribution of trajectory slopes.}
\label{hist_zdavg}
\end{figure}

\begin{figure}
\includegraphics[width=0.99\columnwidth]{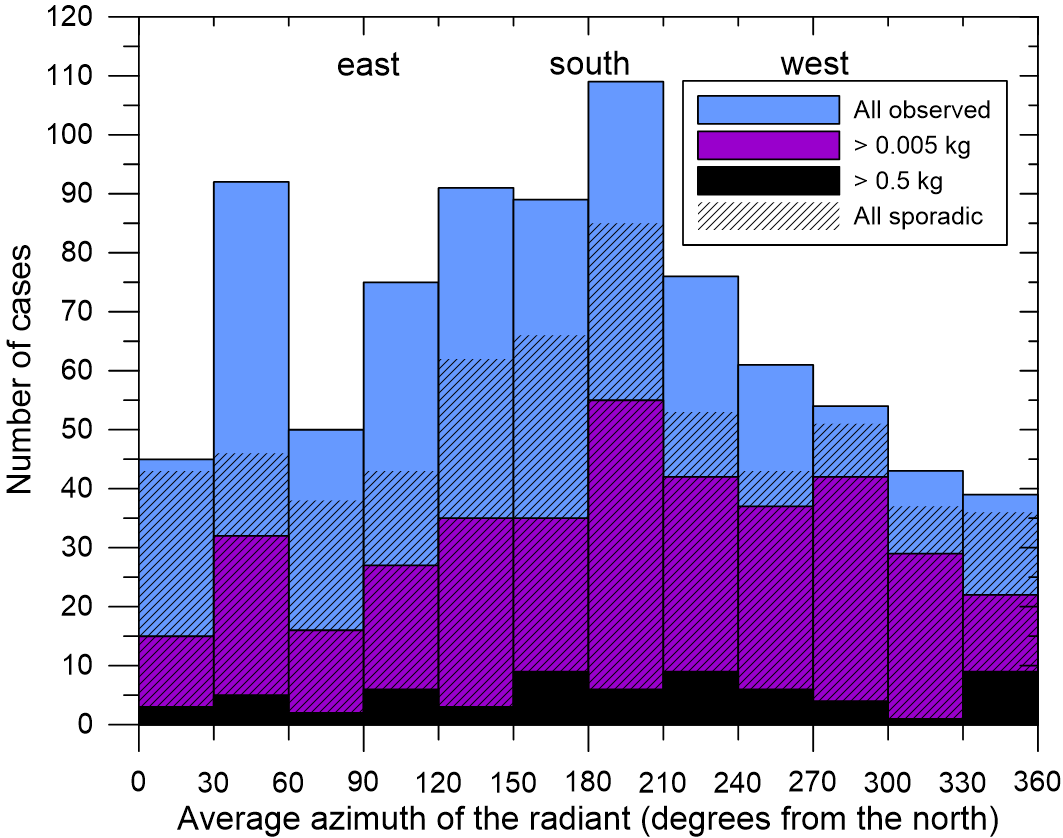}
\caption{Distribution of trajectory azimuths. Azimuth zero corresponds to the north.}
\label{hist_azavg}
\end{figure}

\begin{figure}
\includegraphics[width=\columnwidth]{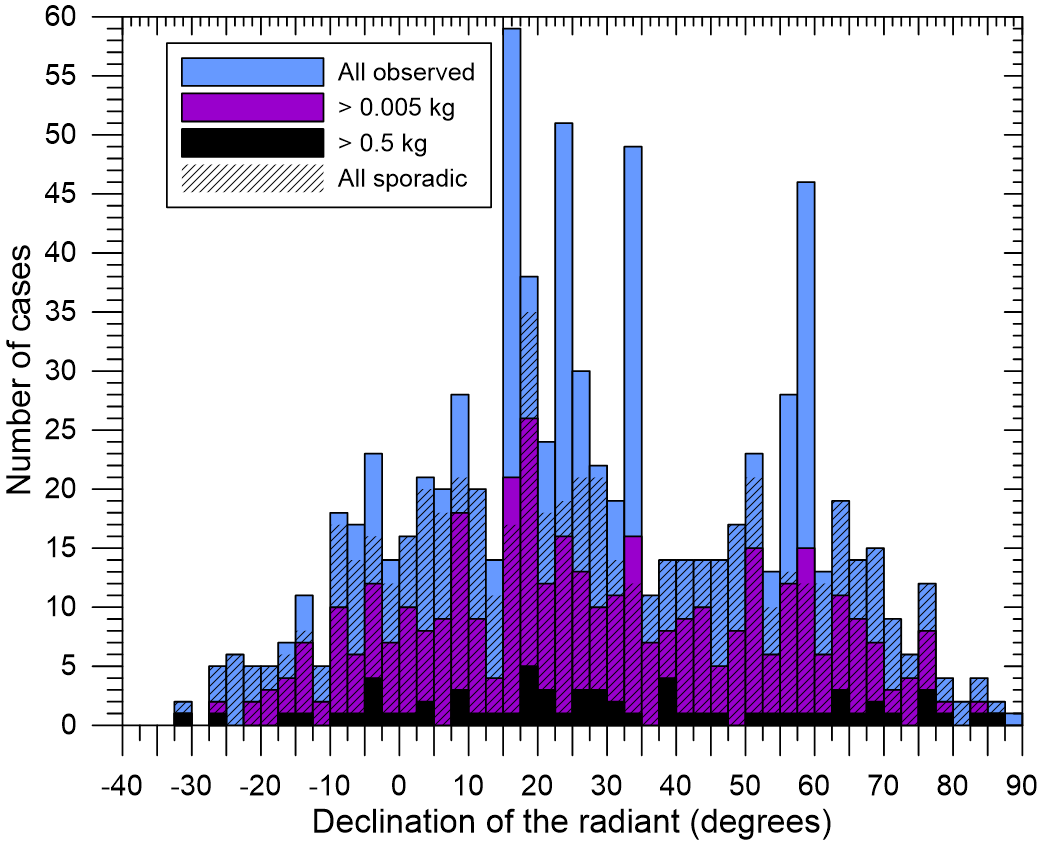}
\caption{Distribution of  apparent declinations of the radiant (J2000.0).}
\label{hist_dec}
\end{figure}

\begin{figure}
\includegraphics[width=\columnwidth]{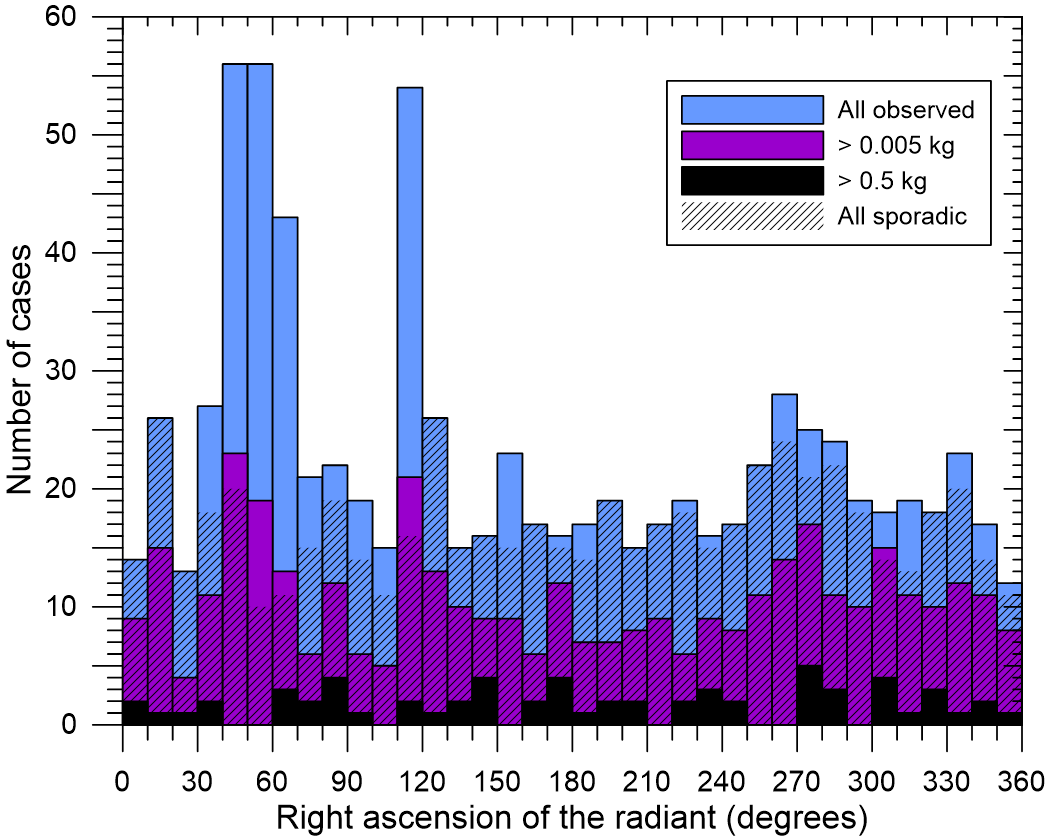}
\caption{Distribution of  apparent right ascensions of the radiant (J2000.0).}
\label{hist_ra}
\end{figure}

\begin{figure}
\includegraphics[width=\columnwidth]{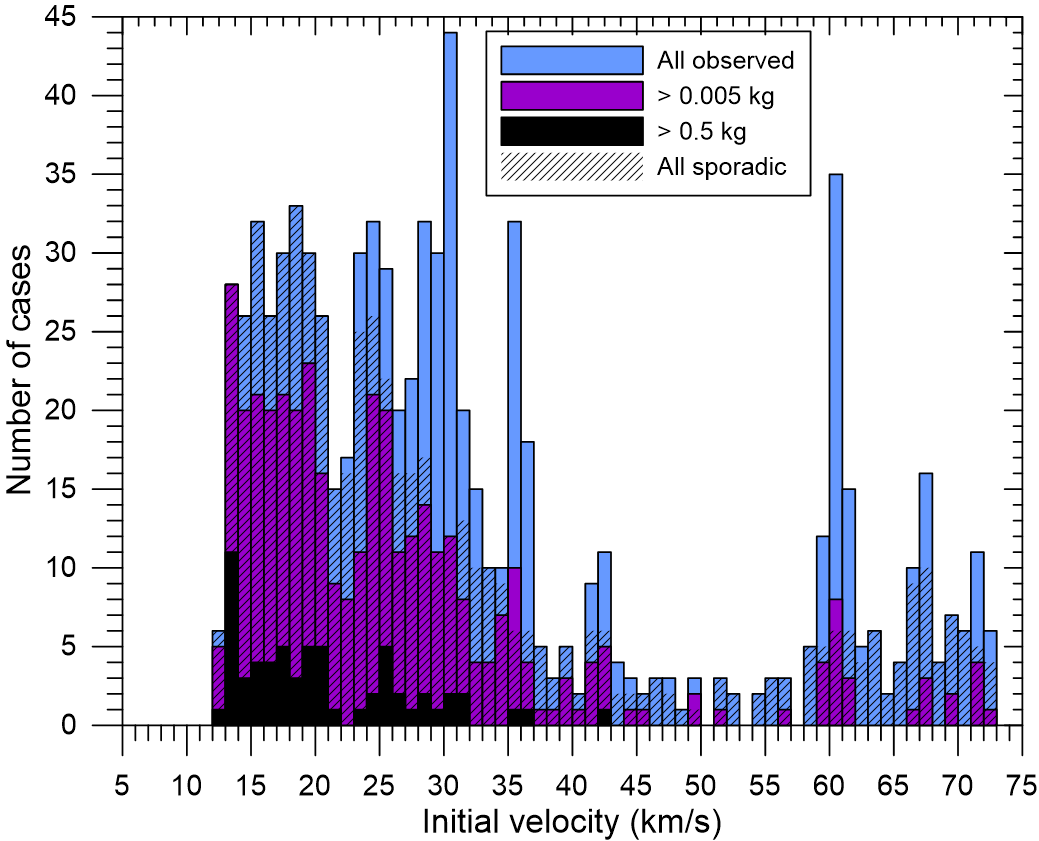}
\caption{Distribution of initial velocities.}
\label{hist_vinf}
\end{figure}

\begin{figure}
\includegraphics[width=\columnwidth]{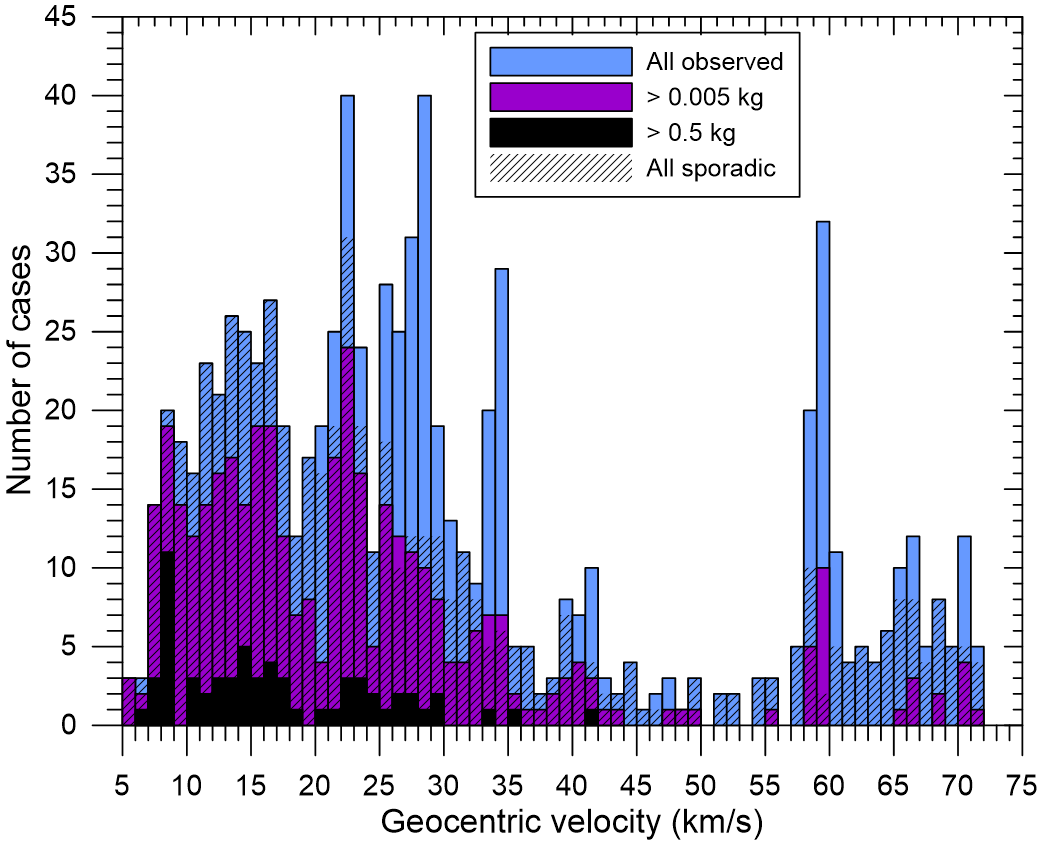}
\caption{Distribution of geocentric velocities.}
\label{hist_vgeo}
\end{figure}

\begin{figure}
\includegraphics[width=\columnwidth]{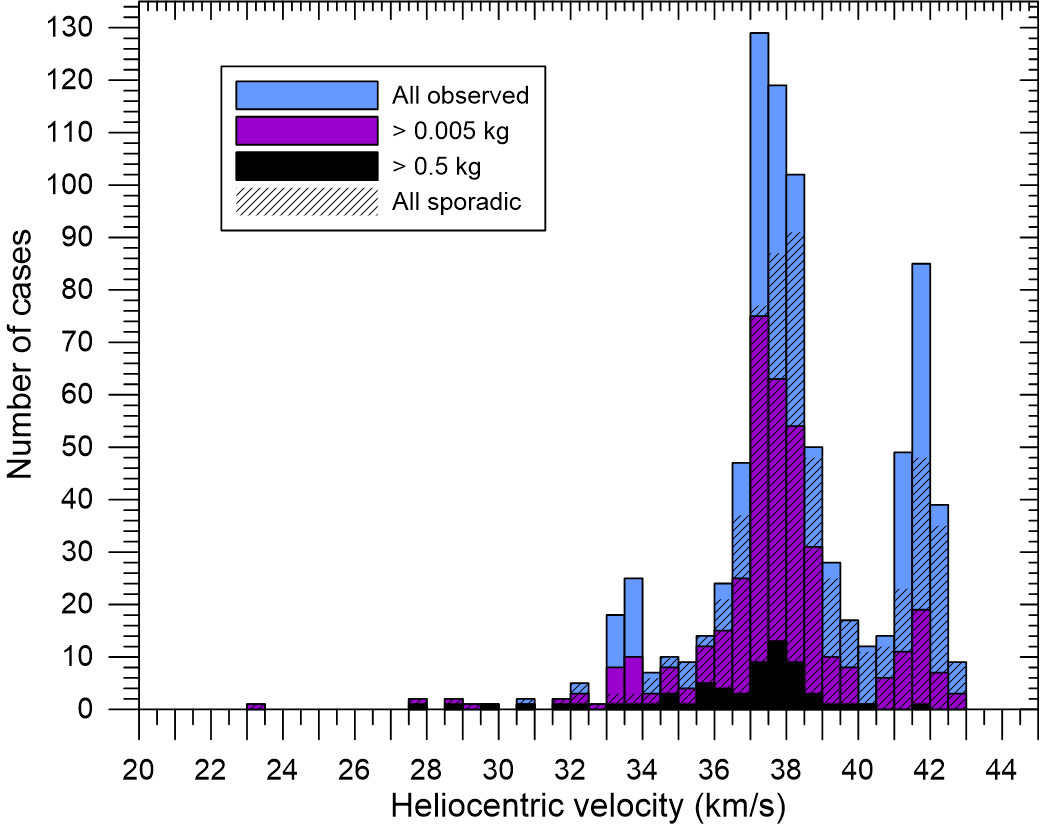}
\caption{Distribution of heliocentric velocities.}
\label{hist_vhelio}
\end{figure}

\begin{figure}
\includegraphics[width=\columnwidth]{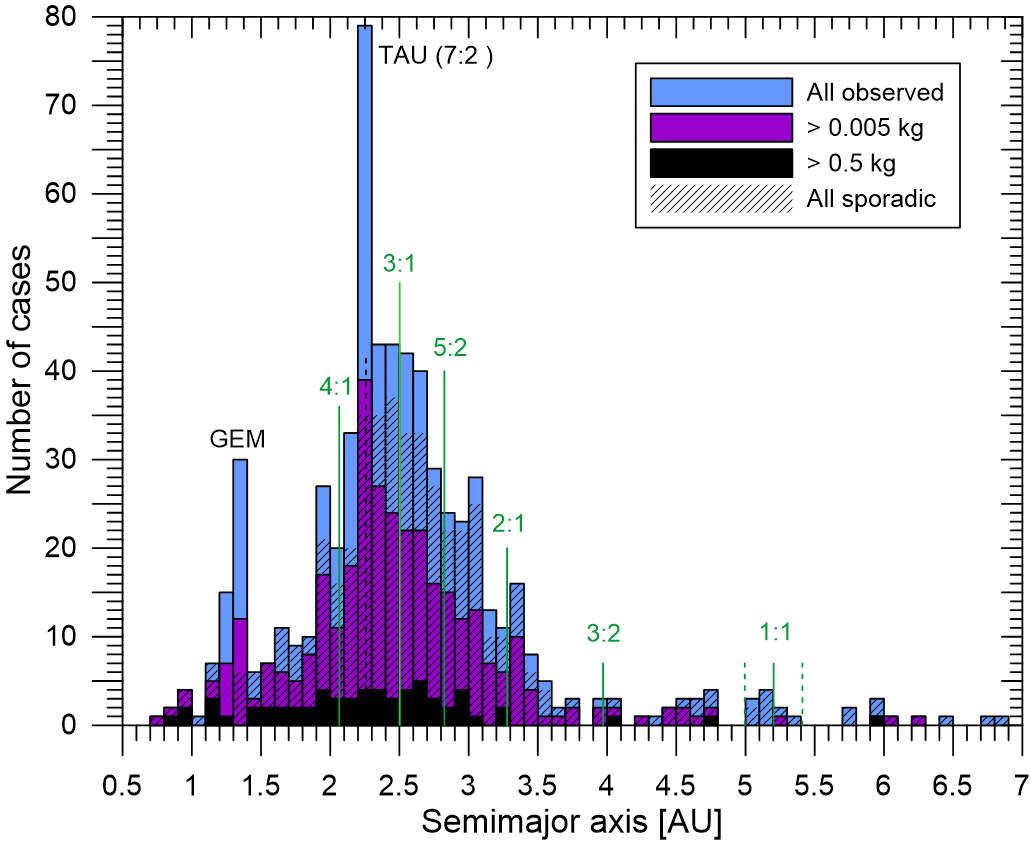}
\caption{Distribution of semimajor axes of short-period meteoroids with $a<7$ AU. The positions of major mean motion resonances
with Jupiter are indicated. Peaks caused by meteor showers Geminids and Taurids are marked.}
\label{hist_axis}
\end{figure}

\begin{figure}
\includegraphics[width=0.95\columnwidth]{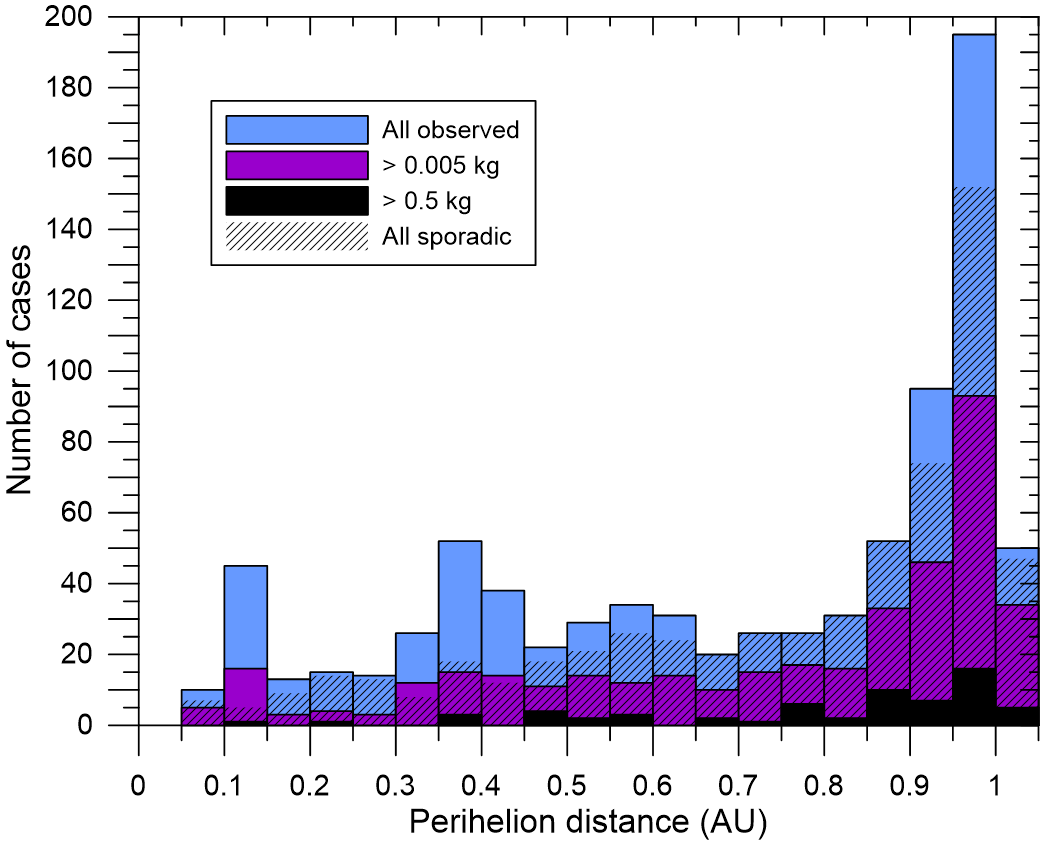}
\caption{Distribution of perihelion distances.}
\label{hist_perihel}
\end{figure}

\begin{figure}
\includegraphics[width=0.95\columnwidth]{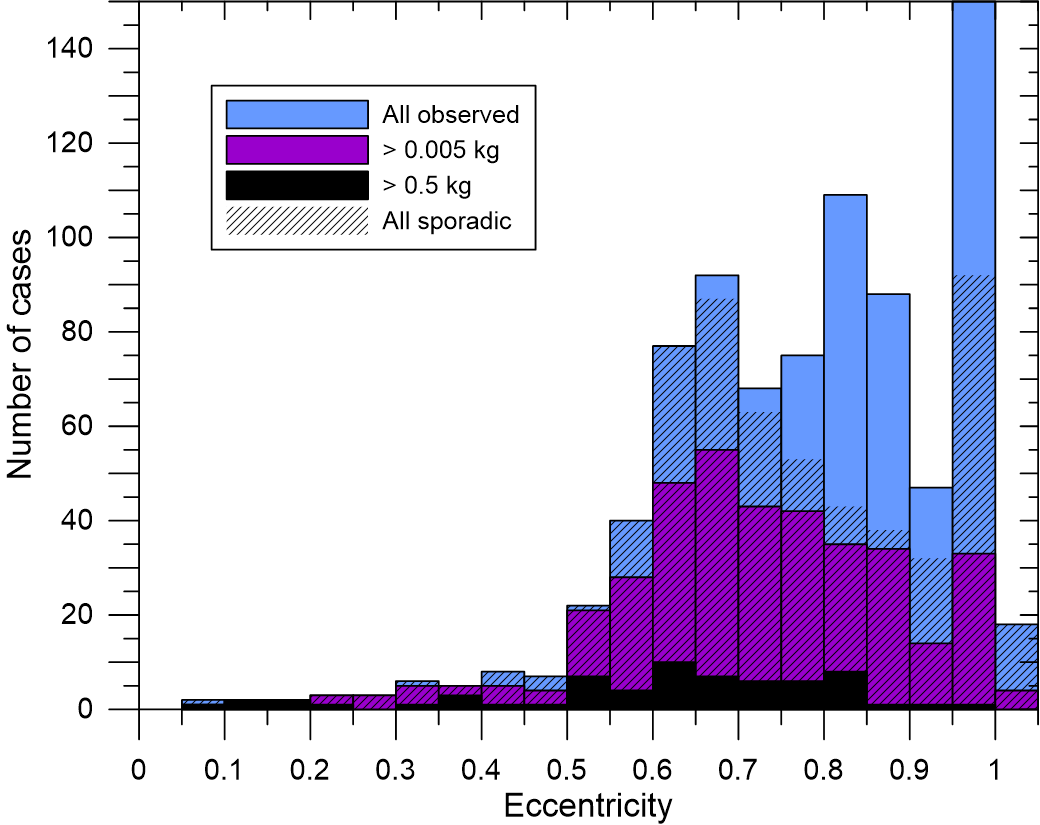}
\caption{Distribution of orbital eccentricities.}
\label{hist_eccl}
\end{figure}

\begin{figure}
\includegraphics[width=\columnwidth]{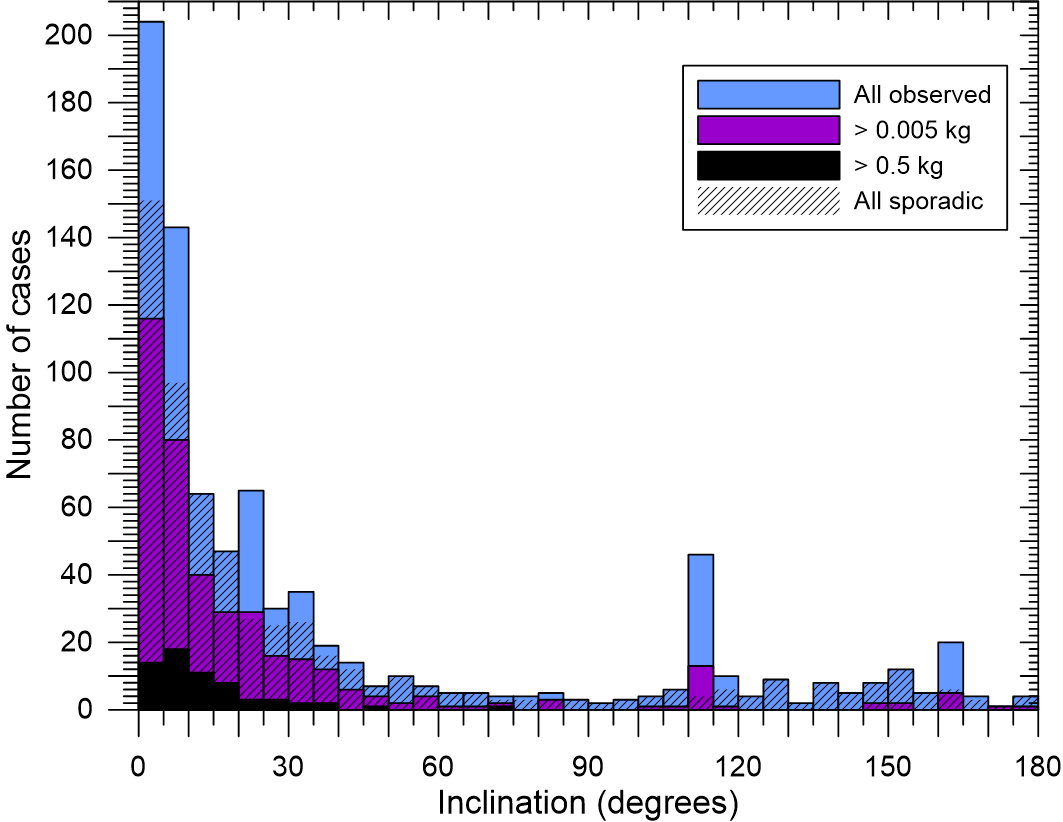}
\caption{Distribution of orbital inclinations.}
\label{hist_incl}
\end{figure} 

\begin{figure}
\includegraphics[width=\columnwidth]{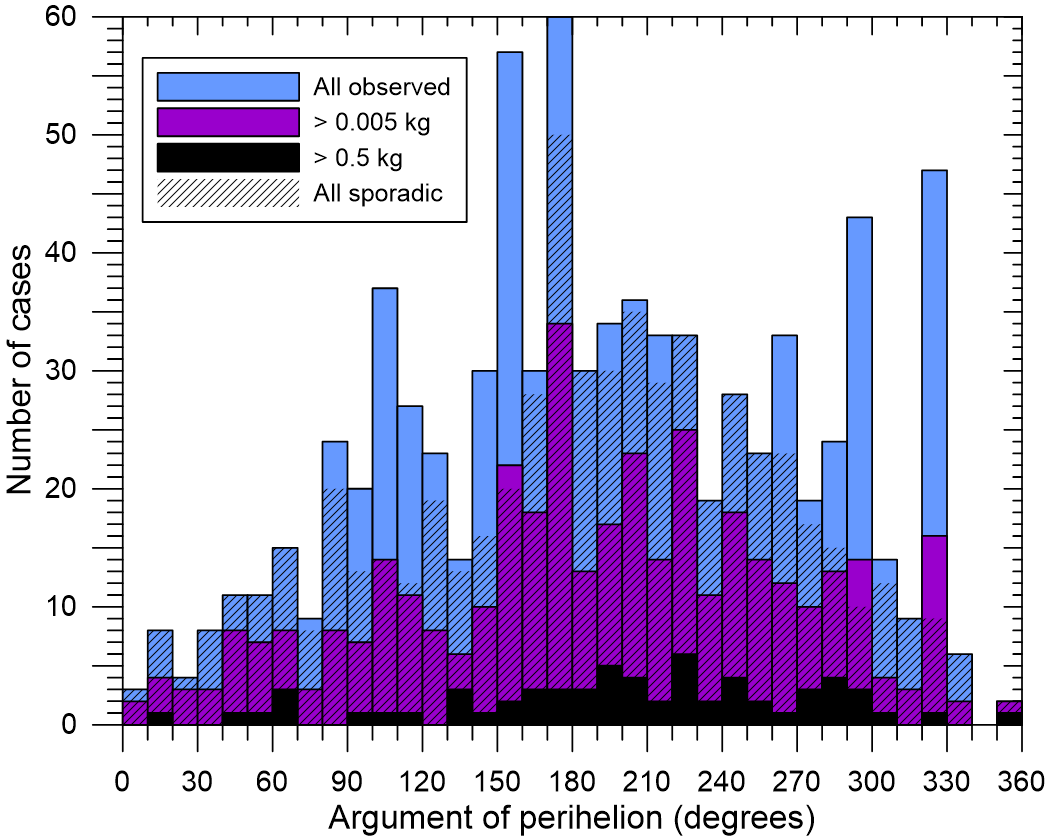}
\caption{Distribution of arguments of perihelia.}
\label{hist_omega}
\end{figure}

\begin{figure}
\includegraphics[width=\columnwidth]{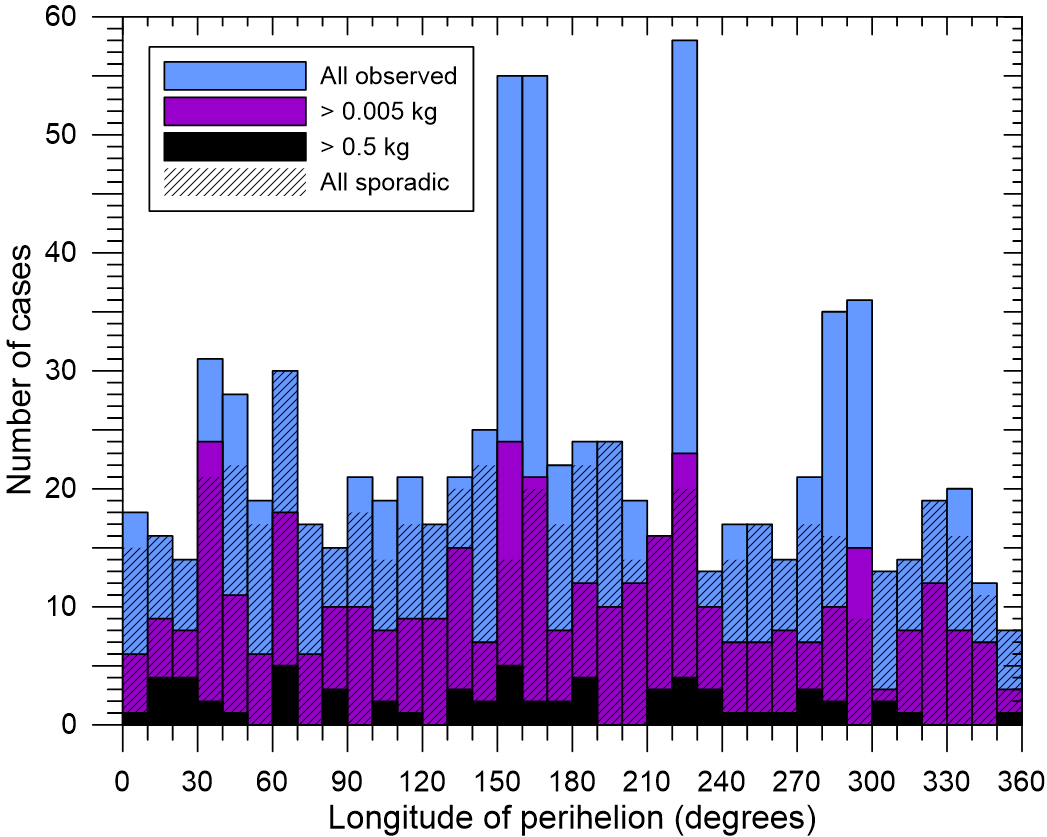}
\caption{Distribution of longitudes of perihelia.}
\label{hist_lperi}
\end{figure}

\begin{figure}
\includegraphics[width=\columnwidth]{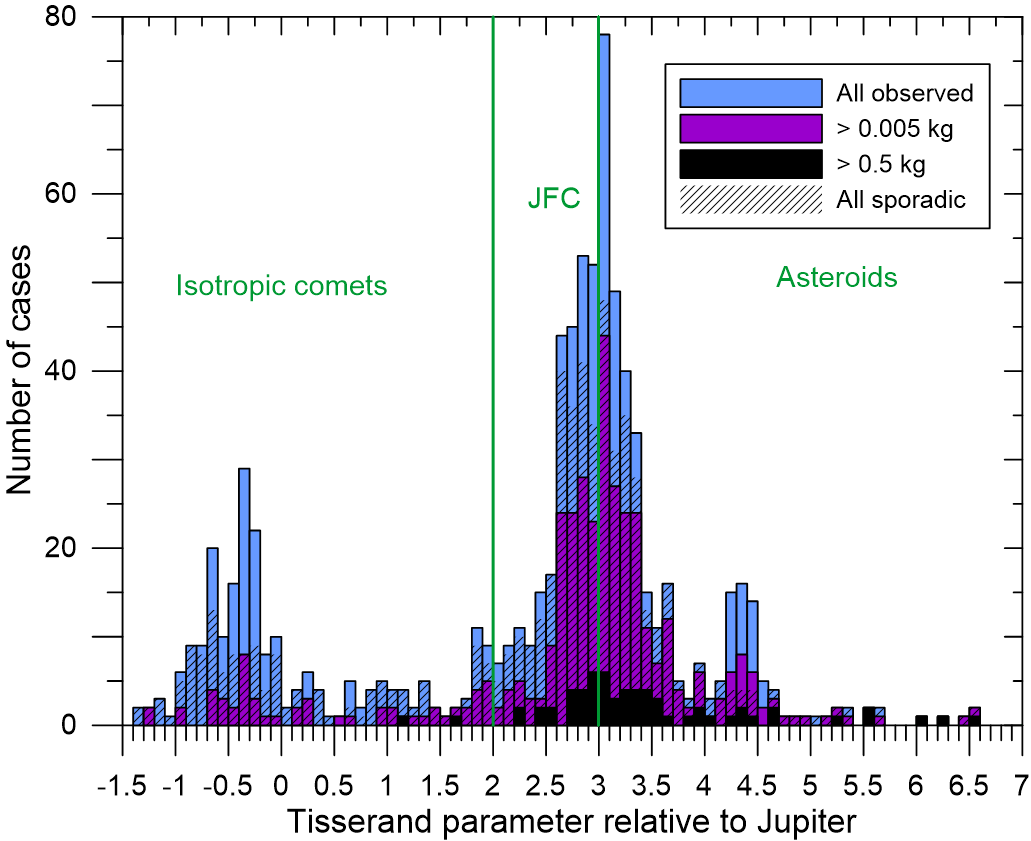}
\caption{Distribution of Tisserand parameters relative to Jupiter. The division into three orbital classes is marked (JFC = Jupiter-family comets).}
\label{hist_Tiss}
\end{figure}

\begin{figure}
\includegraphics[width=\columnwidth]{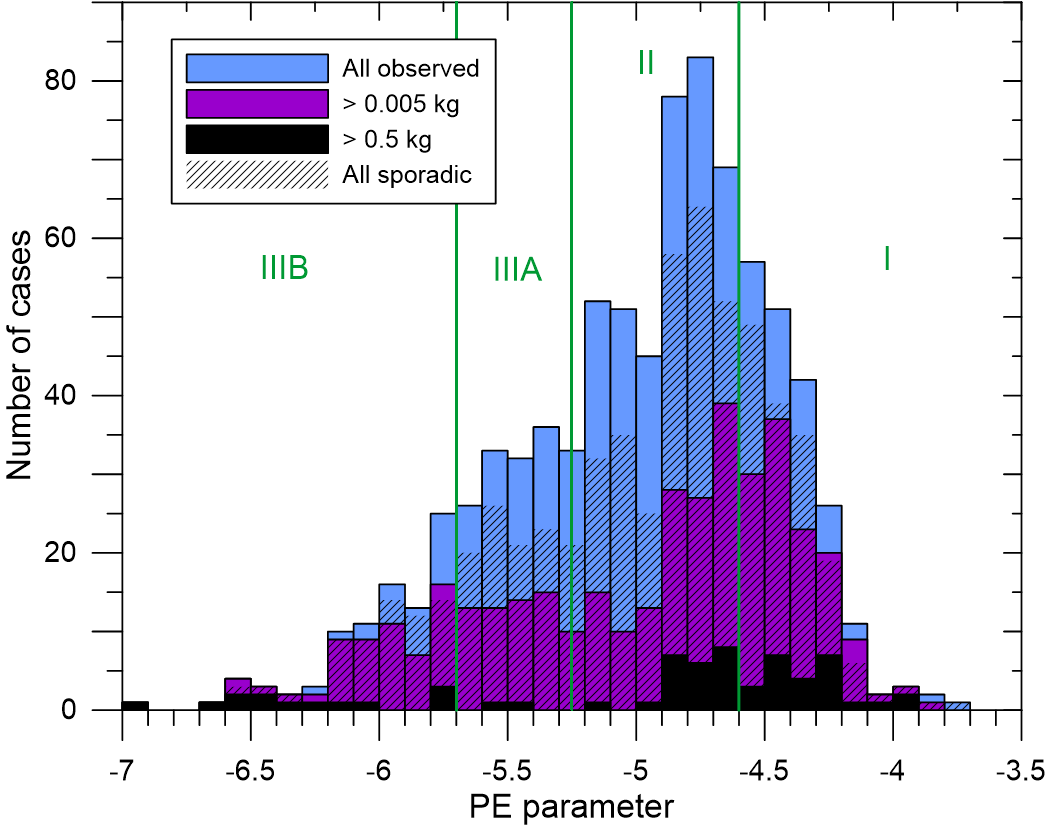}
\caption{Distribution of $P_E$ parameters. The division into four fireball types is marked.}
\label{hist_PE}
\end{figure}

\begin{figure}
\includegraphics[width=\columnwidth]{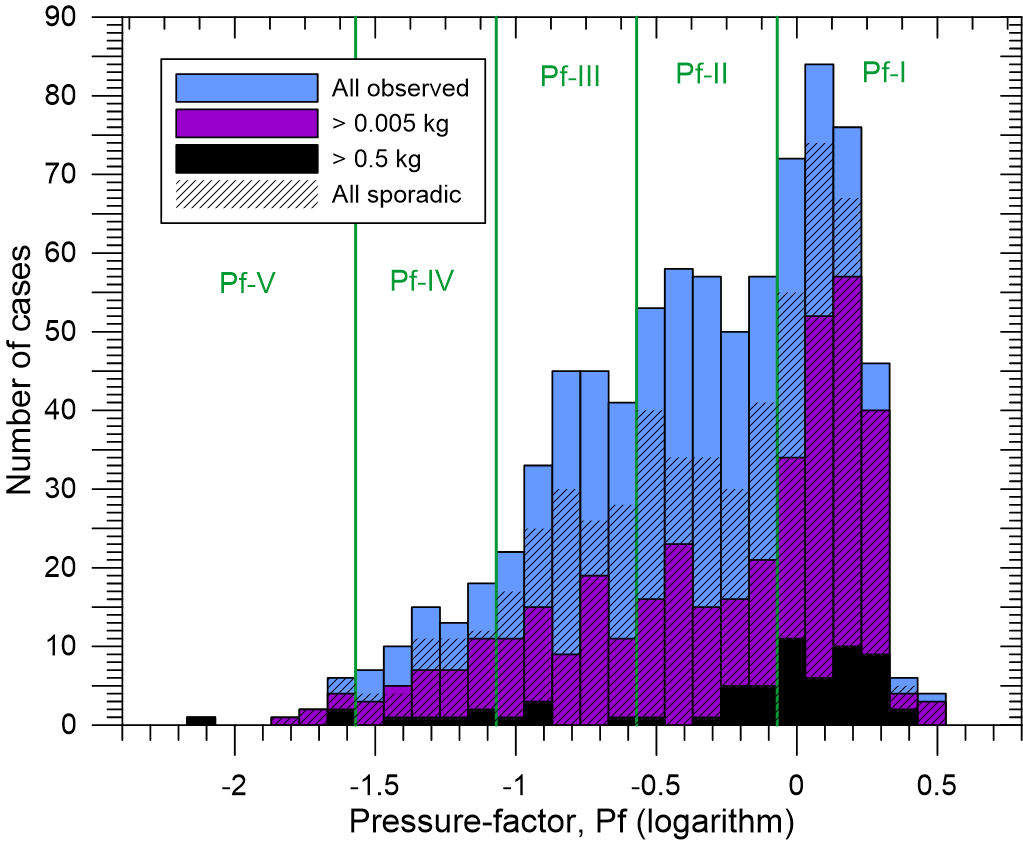}
\caption{Distribution of pressure factors, $P\!f$. The division into five categories of ablation ability is marked.}
\label{hist_Pf}
\end{figure}

\end{appendix}

\end{document}